\documentclass[12pt]{aastex}
\usepackage{natbib}
\shorttitle{Adiabatic horseshoe drag}
\shortauthors{F.\ S.\ Masset \& J. Casoli}
\begin{document}
\title{On the horseshoe drag of a low-mass planet. II Migration in adiabatic disks}
\author{F.\ S.\ Masset\altaffilmark{1,2}} 
\affil{Laboratoire AIM,
  CEA/DSM - CNRS - Universit\'e Paris Diderot, 
  Irfu/Service d'Astrophysique, B\^at. 709, 
  CEA/Saclay, 91191 Gif-sur-Yvette, France}

\altaffiltext{1}{Also at ICF-UNAM, Av. Universidad s/n,
Cuernavaca, Morelos, C.P. 62210, M\'exico} \email{frederic.masset@cea.fr}
\and 
\author{J. Casoli}
\affil{Laboratoire AIM,
  CEA/DSM - CNRS - Universit\'e Paris Diderot, 
  Irfu/Service d'Astrophysique, B\^at. 709, 
  CEA/Saclay, 91191 Gif-sur-Yvette, France}
\email{jules.casoli@cea.fr}
\altaffiltext{2}{Send offprint requests to frederic.masset@cea.fr}
\begin{abstract}
  We evaluate the horseshoe drag exerted on a low-mass planet embedded
  in a gaseous disk, assuming the disk's flow in the coorbital region
  to be adiabatic. We restrict this analysis to the case of a planet
  on a circular orbit, and we assume a steady flow in the corotating
  frame. We also assume that the corotational flow upstream of the
  U-turns is unperturbed, so that we discard saturation effects. In
  addition to the classical expression for the horseshoe drag in
  barotropic disks, which features the vortensity gradient across
  corotation, we find an additional term which scales with the entropy
  gradient, and whose amplitude depends on the perturbed pressure at
  the stagnation point of the horseshoe separatrices.  This additional
  torque is exerted by evanescent waves launched at the horseshoe
  separatrices, as a consequence of an asymmetry of the horseshoe
  region. It has a steep dependence on the potential's softening
  length, suggesting that the effect can be extremely strong in the
  three dimensional case. We describe the main properties of the
  coorbital region (the production of vortensity during the U-turns,
  the appearance of vorticity sheets at the downstream separatrices,
  and the pressure response), and we give torque expressions suitable
  to this regime of migration. Side results include a weak, negative
  feed back on migration, due to the dependence of the location of the
  stagnation point on the migration rate, and a mild enhancement of
  the vortensity related torque at large entropy gradient.
\end{abstract}
\section{Introduction\label{sec:intro}}
Planetary migration is the process by which forming planets undergo
significant variations of their semi-major axis, as a result of the
tidal interaction with the protoplanetary disk. The migration of low
mass planets ($M_p\lesssim 10-15$~$M_\oplus$), called type~I
migration, has long posed a problem for scenarios of planetary
formation. Believed to be systematically inwards and fast
\citep{ww86,w97}, the migration of these planets all the way to the
central object should be much shorter than the disk's lifetime. This
raised the issue of why many planetary systems, including our own,
harbor planets that are not very close to their star, and this
triggered a lot of theoretical efforts, during the last decade, to
exhibit mechanisms that would halt, or even reverse type~I migration.
Long thought to be isothermal (which noticeably simplified the
analytical studies as well as numerical simulations), the tidal
response of the disk was only recently considered in a more realistic
manner, by including an energy equation.  \citet{mota06} noticed that
the disk's response was significantly different in an optically thin
disk (described in the approximation of the shearing sheet, in which
there is no net torque between the planet and the disk), while
\citet{pm06} observed in high resolution, three dimensional
calculations that increasing the opacity of the disk could halt or
reverse the migration of low-mass planets. This result was
subsequently interpreted, in the adiabatic limit, as a result of the
advection of entropy within the horseshoe region \citep{bm08,pp08},
giving rise to a new term of the so-called horseshoe drag. This term
is of crucial importance to theories of planetary migration since it
may halt or revert migration \citep{bm08, 2008A&A...487L...9K,
  pp08}. While there is clearly a link between the advection of
entropy and the existence of an additional term to the horseshoe drag,
which undoubtedly scales with the radial gradient of entropy
\citep{bm08}, there is not yet a fully self-consistent treatment of
the dynamics of the horseshoe region in an adiabatic disk. Even worse,
it is far from clear where the additional contribution originates. So
far, it has been interpreted as due to overdense and underdense
regions which appear at the horseshoe U-turns as the result of the
advection of entropy in a disk that maintains a pressure
balance. These regions are strikingly apparent in numerical
simulations. They end abruptly, by a contact discontinuity, at the
downstream separatrices \citep[see e.g.][]{bm08}. It is easy to
realize that the contribution arising from the perturbed density in
these regions scales with the entropy gradient, and has the correct
sign (as inferred from numerical simulations). Nevertheless, a simple
attempt to perform a full horseshoe drag calculation by integrating on
the horseshoe streamlines reveals a paradox.  The standard horseshoe
drag calculation consists in integrating over the upstream flow, on
each side of the planet, the mass flow rate inside of the stream tubes,
multiplied by the jump of angular momentum experienced by the fluid
particles that execute a horseshoe U-turn in these tubes
\citep{wlpi91}. The net result corresponds to the rate of exchange of
angular momentum between the planet and the gas of the coorbital
region. On the upstream side of the horseshoe U-turns, the mass flow
rate is that of the unperturbed disk, and it cannot depend on the
equation of state of the gas. Therefore, if the torque due to a given
stream tube depends on the equation of state, this should be due to
the fact that the jump of angular momentum during the horseshoe U-turn
depends on the equation of state. For the sake of definiteness,
consider the side of the horseshoe region where we expect the material
to become underdense, and assume that the disk has a vanishing vortensity gradient. 
In order to decrease the density downstream of
the horseshoe U-turns, the flow has essentially two possibilities:
streamlines can move apart (hence horseshoe U-turns are not radially
symmetric), or the velocity on the streamlines can differ from the
Keplerian velocity, and in this case be larger than the Keplerian
velocity in the corotating frame, so as to lead to an expansion of the
material. In other words, a fluid particle originally on a streamline
at $-x=r-r_p$ ($r$ being its orbital radius and $r_p$ that of the
planet) is not mapped to $x$ but to $x+\xi$, and its velocity in the
corotating frame is not $V=2A(x+\xi)$ but $V=2A(x+\xi)+v$ (here
$A=(1/2)r\partial_r\Omega/\partial r$ is the first Oort's constant,
that quantifies the Keplerian shear at the planet's orbit).  The
perturbed elongation $\xi$ and perturbed velocity $v$ adjust so as to
create an underdense region downstream of the horseshoe U-turns.  A
simple relationship between $\xi$ and $v$ can be found using the
conservation of the Bernoulli invariant in the corotating frame. For
the sake of simplicity, we assume that here there is no pressure gradient,
hence the variation of enthalpy during the horseshoe U-turn cancels
out for an adiabatic flow. The conservation of the Bernoulli constant
therefore reduces to the conservation of the Jacobi constant $J$,
which reads $J=V^2/2+2\Omega Ax^2$. To lowest order in $v$ and $\xi$,
the conservation of the Jacobi constant yields $2B\xi + v=0$ (where
$B=\Omega+A$ is the second Oort's constant). The jump of angular
momentum of fluid elements is, to lowest order, $\Delta j=2r_pBx +
2r_pB(x+\xi) + r_pv$, hence it is $\Delta j=4r_pBx$, the very same
value of the jump that is used in classical, barotropic estimates of
the horseshoe drag. It seems therefore that there is no room, in a
classical horseshoe drag formulation, for a dependence on the
thermodynamics of the gas, whereas such dependence is evidenced by
numerical simulations\footnote{Although our simple example is limited
  to a case with no pressure gradient and no vortensity gradient, we also expect an adiabatic
  torque excess in this case.}.  In this work we address this
apparent contradiction. We lay out our assumptions, and define our
notation and conventions in section~\ref{sec:prerequisite}. We
introduce a formalism that leads to a rigorous horseshoe drag
expression in section~\ref{sec:hors-drag-expr}. This expression
features a dependence on the entropy gradient, as expected. We call
the corresponding term the adiabatic torque excess. We compare this
adiabatic torque excess to results of numerical simulations in
section~\ref{sec:comp-numer-simul}.  We present the features of
the horseshoe region that most differ from the barotropic case in
section~\ref{sec:prop-coorb-regi}. These are essentially the
appearance of vorticity sheets at the downstream separatrices and a
mild production of vortensity all over the downstream sides of the
horseshoe region.  We also work out the pressure and density response
over the coorbital region, and identify the term that is related to
the adiabatic torque excess. In section~\ref{sec:interpr-an-intr}, we
give a simple interpretation of the origin of the adiabatic excess,
and we provide suitable torque expressions in
section~\ref{sec:suit-torq-expr}. We then discuss additional issues in
section~\ref{sec:discussion}, and draw our conclusions in
section~\ref{sec:conclusions}.

\section{Prerequisite}
\label{sec:prerequisite}
\subsection{Notation}
\label{sec:notation}

We consider a planet of mass~$M_p$ orbiting a star of mass~$M_*$, on a
fixed circular orbit of radius~$a$ and of angular
frequency~$\Omega_p$. The planet is immersed in a gaseous
protoplanetary disk, such that the planetary orbit is coplanar with
the disk and prograde. We use $\Sigma$ to denote the disk's surface
density, $P$ to denote the vertically integrated pressure, and $T$ to
denote the vertically averaged temperature. We assume that the disk's
gas follows the ideal gas law, which reads:
\begin{equation}
  \label{eq:1}
  P=\frac{{\cal R}\Sigma T}{\mu},
\end{equation}
in which ${\cal R}$ is the ideal gas constant and $\mu$ is the mean
molecular weight. 
The disk isothermal
sound speed is therefore:
\begin{equation}
  \label{eq:2}
  c_s^{\rm iso}=\sqrt{\frac{{\cal R}T}{\mu}},
\end{equation}
while adiabatic sound waves propagate at the speed:
\begin{equation}
  \label{eq:3}
  c_s=\sqrt{\frac{\gamma{\cal R}T}{\mu}},
\end{equation}
where $\gamma$ is the usual adiabatic index.
We shall consider the gas entropy in two flavors. The first of those,
which we denote with a lower $s$, is oftentimes used in the
astrophysical literature owing to its simplicity, as its reads:
\begin{equation}
  \label{eq:4}
  s = \frac{P}{\Sigma^\gamma}.
\end{equation}
Nevertheless, we shall also need an expression for the entropy that
is compliant with the original definition of entropy (the entropy
exchanged is the heat exchanged divided by the temperature). We denote
the latter with an upper $S$:
\begin{equation}
  \label{eq:5}
  S = \frac{{\cal
      R}}{(\gamma-1)\mu}\log\left(\frac{P}{\Sigma^\gamma}\right)
=  \frac{{\cal
      R}}{(\gamma-1)\mu}\log s.
\end{equation}
Another thermodynamics variable that we shall need is the enthalpy
$\eta$,
which reads for an ideal gas:
\begin{equation}
  \label{eq:6}
  \eta = \frac{\cal R}{\mu}\cdot\frac{\gamma}{\gamma-1}T.
\end{equation}
In what follows we shall assume, for the sake of simplicity, that
${\cal R}/\mu = 1$, which amounts to changing the units of
temperature, entropy and enthalpy, without loss of generality.

We note $T_0$ the temperature of the unperturbed disk, which we
assume to be uniform.

We identify a location in the disk by its distance $r$ to the star and
its azimuth $\phi$ with respect to the planet. The gas has a radial
velocity $v_r$ and an azimuthal velocity $v_\phi$ in the frame corotating
with the planet (hence its angular frequency in an inertial frame is
$\Omega=v_\phi/r+\Omega_p$). The radius of corotation $r_c$ is the
location in the unperturbed disk where the material has same angular
frequency as the planet: $\Omega(r_c)=\Omega_p$. We will make use of
the distance~$x$ to corotation: $x=r-r_c$. We also consider the disk
pressure scaleheight $H=c_s/\Omega$, and the aspect ratio $h=H/r$.

The gravitational potential exerted on the disk can be decomposed as
the sum of the stellar potential $\Phi_*=-GM_*/r$, of the planetary
potential $\Phi_p$ and of the indirect potential $\Phi_i$. The
expression of these last two terms are respectively:
\begin{equation}
  \label{eq:7}
  \Phi_p = -\frac{GM_p}{(r^2-2ra\cos\phi+a^2+\epsilon^2)^{1/2}},
\end{equation}
where $\epsilon$ is the softening length of the planetary potential,
and
\begin{equation}
  \label{eq:8}
  \Phi_i = \frac{GM_p}{a^2}r\cos\phi=qa\Omega_p^2r\cos\phi,
\end{equation}
where $q=M_p/M_*$.

Using the notation of \cite{bm08}, we define the gradient of vortensity
across corotation as:
\begin{equation}
  \label{eq:9}
  {\cal V}=\frac{d\log\left(\Sigma/\omega\right)}{d\log r}=\frac
  32+\frac{\Sigma_c}{r_c}
\left.\frac{d\Sigma}{dr}\right|_{r_c},
\end{equation}
where $\omega=(1/r)\partial_r(r^2\Omega)$ is the vertical component of
the flow vorticity, and where $\Sigma_c$ is the unperturbed surface
density at the corotation radius.
 Note that instead of the vortensity properly speaking, which
is $\omega/\Sigma$, we shall oftentimes consider its inverse,
$\Sigma/\omega$. One reason for this choice is that, if we consider
that the flow is that of the unperturbed disk, the perturbations of
the inverse of the vortensity can directly be converted into
perturbations of density, whose impact on the torque is
straightforward.
Finally, we define the gradient of entropy across corotation as:
\begin{equation}
  \label{eq:10}
  {\cal S}=\frac 1\gamma\frac{d\log s_0}{d\log r}=\frac{\gamma-1}{\gamma}r\partial_rS_0.
\end{equation}
\subsection{Basic equations}

The governing equations of the flow are the equation of continuity, the Euler equation and the energy equation, together with the closure relationship provided by the equation of state.
The equation of continuity reads, in the frame corotating with the planet:
\begin{equation}
  \label{eq:11}
  \partial_t\Sigma+\frac 1r\partial_r(\Sigma rv_r)+\frac
  1r\partial_\phi(\Sigma v_\phi)=0.
\end{equation}
The Euler equations read, respectively in $r$ and $\phi$:
\begin{equation}
  \label{eq:12}
  \partial_tv_r+v_r\partial_rv_r+\frac{v_\phi}{r}\partial_\phi
  v_r-r\Omega_p^2-2\Omega_pv_\phi-\frac{v_\phi^2}{r}
=-\frac{\partial_rP}{\Sigma}-\partial_r\Phi,
\end{equation}
and
\begin{equation}
  \label{eq:13}
  D_tj=-\frac{\partial_\phi p}{\Sigma}-\partial_\phi\Phi,
\end{equation}
where
$D_t\equiv \partial_t+v_r\partial_r+\frac{v_\phi}{r}\partial_\phi$
and $j=r^2\Omega$ is the specific angular momentum.
The equation of state that we adopt is that of ideal gases, which reads $P=\Sigma T$. In that case, the
internal energy density is $e=p/(\gamma-1)$, and the energy equation reads:
\begin{equation}
  \label{eq:14}
  \Sigma D_t\left(\frac e\Sigma\right)=-p\vec\nabla.\vec v.
\end{equation}
This equation does not include source or sink terms of energy, as we assume the flow to be adiabatic.

\subsection{Conventions}
Since different authors represent the horsehoe region in different
ways, we find it useful to refer to properties of this region in a manner independent
of the representation.  Orienting the azimuth
rotation-wise, and assuming that, in the corotating frame, the planet
is located at azimuth $\phi=0$, we say that something occurs in front
of the planet if it occurs at $\phi>0$, and behind or on the rear side
of the planet if it occurs at $\phi<0$. Similarly, we refer to the
upstream part of the horseshoe streamlines as the set of fluid
elements that have not performed yet a horseshoe U-turn, whereas the
downstream part correspond to the set of fluid elements that have
already performed their U-turn.

\subsection{Assumptions}
\label{sec:assumptions}
Our main assumptions are as follows:
\begin{itemize}
\item We assume that there is only one stagnation point in the
  vicinity of the planet. This is usually not the case in barotropic
  situations, in which one has generally two X-stagnation points on the
  corotation, on each side of the planet \citep[see e.g.][or
  paper~I]{mak2006}. In the adiabatic case, nevertheless, we usually
  have only one stagnation point, unless either the entropy gradient is very
  small, or the potential's softening length is very small. We shall therefore assume in what follows that there is no
  ambiguity when we refer to the stagnation point. We will make some
  additional remarks on this assumption in
  section~\ref{sec:topol-flow-locat}.
\item We assume that the initial temperature field is uniform. This
  assumption is very similar to the assumption of global isothermality
  that is required in the isothermal case to carry out a rigorous
  horseshoe drag calculation (see paper~I). This assumption is
  required in order for the quantity that we shall define in
  section~\ref{sec:useful-invariant} to be conserved along a fluid
  element's path. We will mention in
  section~\ref{sec:extens-arbitr-temp} how our results can be
  generalized to the case of an arbitrary temperature profile.
\item Finally, we assume that the flow is in steady state in the frame
  corotating with the planet, and that the flow upstream of the
  horseshoe U-turns is unperturbed. We note that these two assumptions
  are contradictory: reaching a steady state implies that the flow has
  executed many horseshoe librations, hence the upstream flow cannot
  be that of the unperturbed disk. Nevertheless, we restrict
  ourselves, in this analysis, to a study of the unsaturated torque
  value. This implies that we consider the flow on a time scale (i)
  longer than what it takes to execute a horseshoe U-turn, so that the
  flow has reached a steady state in some region of interest enclosing
  the planet, (ii) shorter than the horseshoe libration time, so that
  in the region of interest any fluid element has executed at most one
  horsehoe U-turn.
\end{itemize}

\subsection{A useful invariant}
\label{sec:useful-invariant}
We consider the Jacobi-like Bernoulli constant whose expression is, in steady state:
\begin{equation}
\label{eq:15}
  B_J=\frac{r^2(\Omega-\Omega_p)^2+v_r^2}{2}+\Phi_*+\Phi_p+\Phi_i
-\frac 12r^2\Omega_p^2+\eta.
\end{equation}
Since $B_J$ is conserved along the path of a fluid element, the
quantity defined as:
\begin{equation}
\label{eq:16}
G=B_J-T_0S
\end{equation}
is also conserved in an isentropic flow. Far from the planet, assuming
a purely azimuthal motion, we can write, from Eqs.~(\ref{eq:15})
and~(\ref{eq:16}):
\begin{equation}
\label{eq:17}
\partial_rG=(\Omega-\Omega_p)r\omega+(T-T_0)\partial_rS,
\end{equation}
where we have used the rotational equilibrium, which can be deduced
from Eq.~(\ref{eq:12}) by letting $v_r\equiv 0$, and which reads:
\begin{equation}
\label{eq:18}
\partial_r\phi_*+\partial_r\eta=r\Omega^2+T\partial_rS.
\end{equation}
As a consequence, in the unperturbed flow ($T\equiv T_0$), $G$ is
maximal at corotation, as is $B_J$ in an isothermal disk, and it
admits the following expansion to second order in $x$:
\begin{equation}
\label{eq:19}
G=G_c+2AB x^2,
\end{equation}
where $G_c$ is the value of $G$ at corotation, $A=(1/2)rd\Omega/dr$ is
the first Oort's constant and $B=(1/2r)d(r^2\Omega)/dr$ is the second
Oort's constants, which are to be evaluated at corotation.
\section{A horseshoe drag expression}
\label{sec:hors-drag-expr}
\subsection{Initial formulation}
\label{sec:hors-drag-form}
\begin{figure}
 \centering
 \plotone{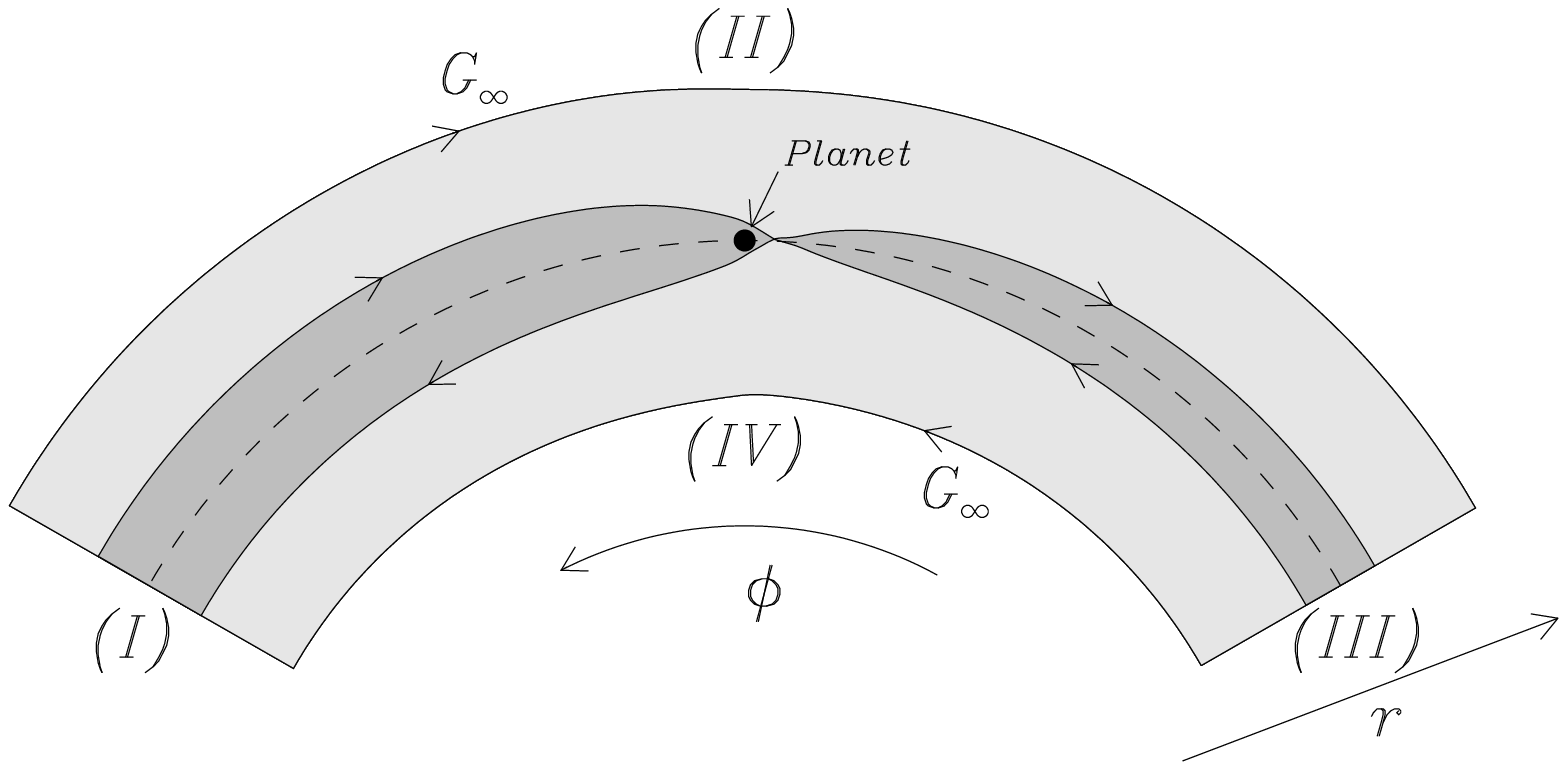}
 \caption{Sketch of the domain of integration (light grey) enclosing the horseshoe region (dark grey).
The dashed line shows the corotation. As can be seen here, there is usually an offset between the planet, and the stagnation
point of the horseshoe region (the point of intersection of the boundaries of the horseshoe region, i.e. the separatrices).
There is also generally an offset between the corotation radius and the orbital radius of the planet, but it is much
smaller.}
 \label{fig:scheme1}
\end{figure}

We consider a domain~$D$ of the disk consisting of an angular sector
centered on the star that encloses the planet, as depicted in
Fig.~\ref{fig:scheme1}. The azimuth of the boundaries~(I) and~(III) is
chosen sufficiently large so that the material that crosses them can
be considered to have a purely circular motion, while the
boundaries~(II) and~(IV) enclose the horseshoe region, and are chosen
much further away from corotation than the horseshoe separatrices, so
that they can be considered as circular. We note that in a real
situation, this last requirement is hindered by the wake, which
corresponds to epicyclic motion excited at Lindblad resonances.
Nevertheless we discard this behavior in the present work, which
amounts to considering an isolated corotation region. We want to avoid
flux of material into the domain~$D$ through the boundaries~(II)
and~(IV), so we adopt streamlines to define them.  These two boundaries
are therefore associated to some value of the~$G$ invariant. For
convenience, we use the same value for $G$ on both sides, and we note
it $G_{\infty}$\footnote{ Since $G=2ABx^2$ to lowest order in $x$, it is certainly
possible to adopt the same value of the $G$ invariant for the outer and inner
boundaries of the integration domain, if these are not too far from corotation.
If they are located at a sizable fraction of the semi-major axis from corotation, however, this may
not longer be possible, and one may have different values of the $G$-invariant
for the inner and outer boundaries. This does not change the generality of the
demonstration presented here.}.

The torque exerted on the planet by the material included in the
domain $D$ reads:
\begin{equation}
  \label{eq:20}
  \Gamma= \int\!\int_D
\Sigma(\partial_\phi\Phi)\,d\phi\,rdr\,,
\end{equation}

We can transform the integrand of Eq.~(\ref{eq:20}) as follows:
\begin{equation}
  \label{eq:21}
  \Sigma(\partial_\phi\Phi)r=-r\partial_\phi
  P-\partial_\phi(v_\phi\Sigma j)-\partial_r(r\Sigma v_r j),
\end{equation}
where we have made use of the assumption of a steady state in the
corotating frame, and where we have used Eqs.~(\ref{eq:11})
and~(\ref{eq:13}).

Since the boundaries of the domain~$D$ are sufficiently far
from the planet that we can neglect radial motions, Eq.~(\ref{eq:20})
can be recast as:
\begin{equation}
  \label{eq:22}
  \Gamma = \left[\int_{r^-_{\infty}}^{r^+_{\infty}}
  [rP+r(\Omega-\Omega_p)\Sigma j]dr\right]_F^R,
\end{equation}
where $r^+_{\infty}$ ($r^-_{\infty}$) is the radius of the outer (inner) streamline
that has $G=G_\infty$.  
The $R$ and $F$ notation at the right bracket of Eq.~(\ref{eq:22}) means that
the integration over $r$ has to be performed respectively at the rear or front
side of the domain~$D$ (boundaries~III and~I in Fig.~\ref{fig:scheme1}).
In Eq.~(\ref{eq:22}), the first term of the
integrand represents the pressure torque exerted on the material
enclosed within the domain~$D$, while the second term represents the
budget of angular momentum brought to this region by advection. Since
the flow is steady in the corotating frame, the angular momentum of
this domain is constant in time and the torque is therefore integrally
transmitted to the planet.  We recognize in the second term of the
integrand of Eq.~(\ref{eq:22}) the classical horseshoe drag expression
\citep{wlpi91,wlpi92,masset01,mp03,mak2006,2009arXiv0901.2265P}, but this equation
also shows the pressure contribution, which has been overlooked in
previous analysis.

\subsection{A simplifying assumption}
\label{sec:simpl-assumpt}
We assume that in the unperturbed flow, far from the planet, lines of
constant entropy and of constant $G$ coincide. This is not exactly
true: in the unperturbed disk the lines of constant entropy are
circles centered on the primary, while, owing to the presence of the
potential's indirect term given by Eq.~(\ref{eq:8}), the iso-$G$ lines
have a variable distance to corotation in the horseshoe region (an
extreme example being provided by the tadpole separatrix). Our
assumption is therefore more appropriate of a situation in which we
discard the indirect term of the potential, or of a shearing sheet
situation. Nevertheless, we shall make hereafter this assumption,
anticipating that our torque expression does not depend sensitively on
it.

This assumption allows to define unambiguously the derivative
$\partial_GS$, which can be obtained from the measure of $S$ and $G$
on two neighboring streamlines, since these variables are intrinsic to
the line. The quantity $\partial_GS$ is therefore also intrinsic to the
streamline, and hence conserved along it.

In the unperturbed disk, this derivative has the following expression:
\begin{equation}
  \label{eq:23}
  \partial_GS=\frac{\partial_rS_0}{4ABx}.
\end{equation}
It therefore diverges at corotation.

Mass conservation yields a relationship which will be useful to
evaluate the production of vortensity during horseshoe U-turns.  We
consider a horseshoe stream tube, of width $\delta G$. The mass flux
across the tube is:
\begin{equation}
  \label{eq:24}
  \dot M=\frac{\delta G}{\partial_rG}r(\Omega-\Omega_p)\Sigma.
\end{equation}
We can transform Eq.~(\ref{eq:17}) to write:
\begin{equation}
  \label{eq:25}
  (1-\delta T\cdot \partial_GS)\partial_rG=(\Omega-\Omega_p)r\omega,
\end{equation}
where $\delta T=T-T_0$. This eventually gives the mass flux:
\begin{equation}
  \label{eq:26}
  \dot M=\delta G\frac\Sigma\omega(1-\delta T\partial_GS).
\end{equation}
We can therefore write the following relationship:
\begin{equation}
  \label{eq:27}
  \left.\frac\Sigma\omega(1-\delta T\partial_GS)\right|_d=\left.\frac\Sigma\omega\right|_u,
\end{equation}
where the subscript $d$ ($u$) respectively denote the quantities downstream
(upstream) of a horseshoe U-turn.

\subsection{An alternate horseshoe expression}

\begin{figure}
  \centering
  \plotone{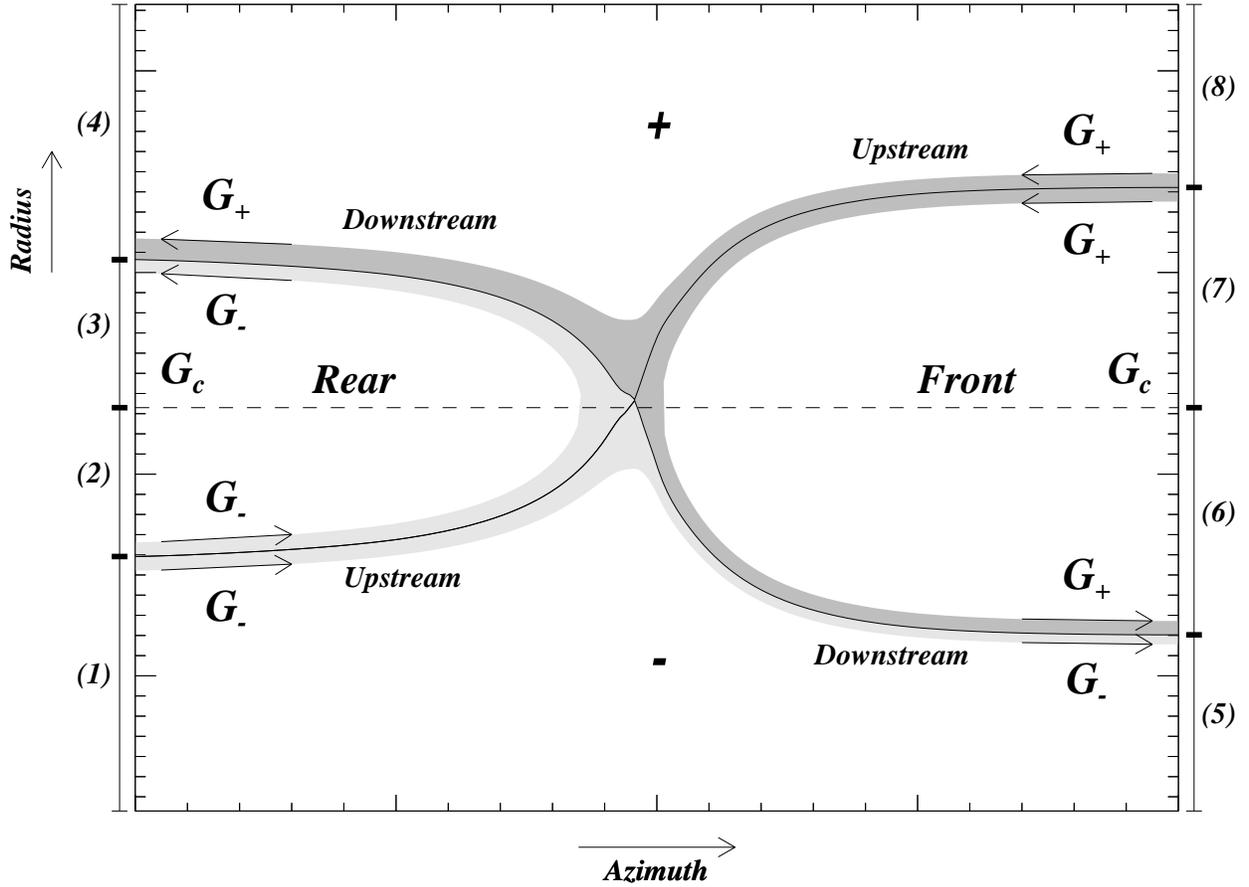}
  \caption{Sketch of the flow topology that shows the value of the $G$
    invariant along the separatrices. The value of the $G$ constant is
    continuous across the upstream separatrices, while it displays a
    jump across the downstream separatrices. The different intervals
    of integration mentioned in the text are shown along the
    $y$-axis. Intervals (1) and (5) are meant to extend up to
    $r^-_\infty$, and intervals (4) and (8) are meant to extend up to
    $r^+_\infty$.}
  \label{fig:quad}
\end{figure}

We now perform a change of variable in the integrals of
Eq.~(\ref{eq:22}). We choose as the new variable of integration the
value of the $G$ invariant. We have therefore to split the intervals
of integration into intervals over which the dependency of $G$ on $r$
is continuous and monotonic.  Anticipating on the results exposed in
the following sections, we note that the value of the $G$ invariant is
not necessarily the same for the front and rear separatrices, as shown
in Fig.~\ref{fig:quad}.  We denote their respective values with $G_+$
and $G_-$.

Using Eq.~(\ref{eq:25}), Eq.~(\ref{eq:22}) is therefore recast as:
\begin{eqnarray}
\label{eq:28}
\Gamma &=& 
\left[
 \int_{r_{\infty}^-}^{r_{\infty}^+}rpdr
+\int_{G_{\infty}}^{G_c}
(1-\delta T\partial_GS)\frac{\Sigma}{\omega}jdG\right.\\ \nonumber
&&\left.+\int_{G_c}^{G_-}
(1-\delta T\partial_GS)\frac{\Sigma}{\omega}jdG
+\int_{G_+}^{G_{\infty}}
(1-\delta T\partial_GS)\frac{\Sigma}{\omega}jdG
\right]_R\\ \nonumber
&&-\left[
\int_{G_{\infty}}^{G_-}
(1-\delta T\partial_GS)\frac{\Sigma}{\omega}jdG
+\int_{G_+}^{G_c}
(1-\delta T\partial_GS)\frac{\Sigma}{\omega}jdG\right.\\ \nonumber
&&\left.+\int_{G_c}^{G_{\infty}}
(1-\delta T\partial_GS)\frac{\Sigma}{\omega}jdG
+  \int_{r_{\infty}^-}^{r_{\infty}^+}rpdr\right]_F\\ \nonumber
\end{eqnarray}
We split the integration on the downstream side of the horseshoe
U-turns in order to account for the possible discontinuity of $G$ at
the separatrices. More precisely, the first integral over $G$ in
Eq.~(\ref{eq:28}) corresponds to an integral over intervals~$(1)$
and~$(2)$ (see Fig.~\ref{fig:quad}), the two following integrals (from
$G_c$ to $G_-$ and $G_+$ to $G_\infty$) correspond respectively to
intervals~$(3)$ and~$(4)$. In a similar manner, on the front side, the
intervals of integration over $G$ are, in order of appearance, $(5)$,
$(6)$ and eventually~$(7)$ and~$(8)$ together. We also note that in
Eq.~(\ref{eq:28}) we have substituted the total pressure $P$ by the
perturbed pressure $p$, since the unperturbed pressure field is
axisymmetric and does not contribute to the pressure torque.

Other comments are in order regarding the transformation of
Eq.~(\ref{eq:22}) into Eq.~(\ref{eq:28}).  The integrand in
Eq.~(\ref{eq:22}) is regular, and although it may display
discontinuities (we shall see that $\Omega$ is discontinuous at the
downstream separatrices, as well as $\Sigma$) it does not contain any
singularity. In contrast, the integrands of Eq.~(\ref{eq:28}) feature
the vorticity, which is singular wherever $\Omega$ is
discontinuous. Since the dependency of $G$ upon $r$ on the different
intervals of integration is regular, the integrals on $G$ must exclude
any singularity (simply because the contribution of the integrand of
Eq.~(\ref{eq:22}) on any vanishingly small interval in $r$ is
vanishingly small, and the same must be true of the equivalent
integral in $G$). As a consequence, any integral of Eq.~(\ref{eq:28})
which has a boundary corresponding to a downstream separatrix must be
understood as excluding the possible singularities at this boundary. For
instance, the second integral over $G$ in Eq.~(\ref{eq:28}) must be understood as:
\begin{equation}
  \label{eq:29}
  \int_{G_c}^{G_-}(1-\delta T\partial_GS)\frac\Sigma\omega jdG\equiv\lim_{\delta G\rightarrow 0^+}
\int_{G_c}^{G_-+\delta G}(1-\delta T\partial_GS)\frac\Sigma\omega jdG.
\end{equation}
We also note that including such singularities would be mathematically
ill-defined anyway, because they lie on the edge of the integration
domain.

Eq.~(\ref{eq:28}) can be further simplified as explained below.  The
specific angular momentum $j$ of a fluid element that has some value
of the $G$ invariant can be written as:
\begin{equation}
  \label{eq:30}
  j=j_0+\delta j,
\end{equation}
where $j_0$ is the specific angular momentum of fluid elements in the
axisymmetric, unperturbed flow which have same value of $G$, and
$\delta j$ is the difference, that we wish to evaluate.  Azimuthally
far from the planet, Eq.~(\ref{eq:16}) can be recast as:
\begin{equation}
  \label{eq:31}
  G=\frac{r^2(\Omega-\Omega_p)^2}{2}+\Phi_*-\frac 12r^2\Omega_p^2+\eta-T_0S,
\end{equation}
where we have neglected the indirect term (as discussed in
section~\ref{sec:simpl-assumpt}). Differentiating Eq.~(\ref{eq:31}),
we can write, to first order in the perturbation:
\begin{equation}
  \label{eq:32}
  0 =r\delta r(\Omega^2-2\Omega\Omega_p)+r^2\delta\Omega(\Omega-\Omega_p)
+\partial_r\Phi_*\delta r+\frac{p}{\Sigma_0}.
\end{equation}
The left hand side of Eq.~(\ref{eq:32}) cancels out because, as stated
above, we consider a fluid element that has same value of $G$ in the
unperturbed flow and in the perturbed one.  Using the rotational
equilibrium of the unperturbed flow, and writing $\delta
P=\partial_rP_0\delta r+p$, we can recast Eq.~(\ref{eq:32}) as:
\begin{equation}
  \label{eq:33}
  0 = (\Omega-\Omega_p)\delta j+\frac{p}{\Sigma_0}.
\end{equation}

We therefore have an expression for $\delta j$ that involves the
perturbation of pressure $p$. Since we consider the variation of
angular momentum of a fluid element for a constant value of $G$, it is
natural to change to $G$ the variable of the integrals of the pressure
torque.  The generic integrand of Eq.~(\ref{eq:28})  therefore becomes:
\begin{eqnarray}
  \label{eq:34}
  \frac\Sigma\omega(1-\delta T\partial_GS)j+\frac{rp}{\partial_rG}&=&
\frac{1}{\partial_rG}[\Sigma r(\Omega-\Omega_p)j_0(G)+\Sigma r(\Omega-\Omega_p)\delta j+rp]\\ 
&=&\frac\Sigma\omega(1-\delta T\partial_GS)j_0(G)+\frac{1}{\partial_rG}\frac{\Sigma_0-\Sigma}{\Sigma_0}rp, \nonumber
\end{eqnarray}
where we have used Eqs.~(\ref{eq:25}) and~(\ref{eq:33}).

The last term of the right hand side of Eq.~(\ref{eq:34}) cancels out
to first order in the perturbation. Furthermore, the parity in $\phi$
of the perturbed density and of the perturbed pressure (which will be
studied in section~\ref{sec:press-dist}) is such that the rear and
front contributions of this term to Eq.~(\ref{eq:28}) cancel
out. Therefore we can discard the pressure torque in
Eq.~(\ref{eq:28}), provided we use the unperturbed value of the
specific angular momentum $j_0$.  Since $G$ has a maximum at
corotation, the notation $j_0(G)$ contains an ambiguity, as one has to
specify if $j_0$ has to be evaluated inside or outside of
corotation. In what follows we use the notation $j_0^\pm$ to remove
this ambiguity, and we drop the $G$ dependency for the sake of
brevity.

The fact that one can neglect the pressure torque and use the value of
the unperturbed angular momentum has been demonstrated in paper~I for
isothermal disks. It is linked (i)~to the fact that the evanescent
pressure waves excited by the perturbation of the flow on the
downstream side of the horseshoe U-turns do not alter the total
horseshoe drag, even though they redistribute the perturbation of
density and therefore change the local torque density, and (ii)~to the fact
that the impact on the torque of the rear-front asymmetry due to the
feed back of the evanescent waves cancels out.

Using the simplification described above, we transform Eq.~(\ref{eq:28}) as
follows:

\begin{eqnarray}
\label{eq:35}
\Gamma &=& 
\left[
\int_{G_{\infty}}^{G_c}
\left.\frac{\Sigma}{\omega}\right|_{0^-}j_0^-dG
+\int_{G_c}^{G_-}
\left.\frac{\Sigma}{\omega}\right|_{0^-}j_0^+dG
+\int_{G_+}^{G_{\infty}}
\left.\frac{\Sigma}{\omega}\right|_{0^+}j_0^+dG
\right]_R\\ \nonumber
&&-\left[
\int_{G_{\infty}}^{G_-}
\left.\frac{\Sigma}{\omega}\right|_{0^-}j_0^-dG
+\int_{G_+}^{G_c}
\left.\frac{\Sigma}{\omega}\right|_{0^+}j_0^-dG
+\int_{G_c}^{G_{\infty}}
\left.\frac{\Sigma}{\omega}\right|_{0^+}j_0^+dG
\right]_F.\\ \nonumber
\end{eqnarray}

Apart from discarding the pressure contribution and changing $j$ into
$j_0$ in Eq.~(\ref{eq:35}), we have introduced an additional
simplification by making use of Eq.~(\ref{eq:27}), and by referring to
the unperturbed, upstream value of $\Sigma/\omega(1-\delta
T\partial_GS)$. We explicitly specify by the use of the index
$0^\pm$ whether the unperturbed vortensity must be considered outside
or inside corotation.

We note (see Fig.~\ref{fig:quad}) that the
integrals over the intervals~(1) and~(5), on the one hand, and over the
intervals~(4) and~(8) on the other hand, exactly cancel each other, hence
the torque expression can further be reduced to:

\begin{equation}
  \label{eq:36}
  \Gamma=
-\int_{G_-}^{G_c}\left.\frac{\Sigma}{\omega}\right|_{0^-}\Delta j_0dG
+\int_{G_+}^{G_c}\left.\frac{\Sigma}{\omega}\right|_{0^+}\Delta j_0dG,
\end{equation}
where $\Delta j_0=j_0^+-j_0^-$. The corotation torque has therefore a
very similar expression to the usual (barotropic) case, when
integrated over a Bernoulli-like variable \citep{mp03}, except that
the domain of integration does not have the same limits on the front
side and on the rear side. Denoting by $G_s$ the arithmetic mean of the
neighboring values $G_-$ and $G_+$, we can transform Eq.~(\ref{eq:36})
into:
\begin{equation}
  \label{eq:37}
  \Gamma \approx \int_{G_s}^{G_c}
\left(
\left.\frac{\Sigma}{\omega}\right|_{0^+}
-
\left.\frac{\Sigma}{\omega}\right|_{0^-}
\right)\Delta j_0dG
-\frac{\Sigma_0}{2B}\Delta j_0^s\Delta G,
\end{equation}
where $\Delta j_0^s$ is the value of $\Delta j_0$ for $G=G_s$, and
where $\Delta G=G_+-G_-$. The left term of the R.H.S. of
Eq.~(\ref{eq:37}) is the standard horseshoe drag, 
which scales with the vortensity gradient,
and the right term, which we hereafter denote with $\Gamma_1$,
is an additional contribution which arises from relaxing the
barotropic hypothesis. Evaluating this contribution therefore amounts
to evaluating $\Delta G$. This is the purpose of the following
section.

\subsection{Discontinuities at the stagnation point}
\label{sec:disc-at-stagn}
The discontinuity in $G$ that we wish to evaluate arises from the
presence, at the downstream separatrices, of a contact
discontinuity (see Fig.~\ref{fig:quad}). Formally, an inspection of Eq.~(\ref{eq:15})
and~(\ref{eq:16}) reveals that this discontinuity can come, far from
the planet, from a discontinuity in the enthalpy, in the entropy, and
in the azimuthal velocity (the other terms being continuous across the
separatrices). The situation becomes simpler at the stagnation point,
where the velocity vanishes and where the discontinuity in $G$ is
entirely accounted for by a discontinuity of entropy and enthalpy.  We
have therefore:
\begin{equation}
  \label{eq:38}
  \Delta G = \eta_+-\eta_--T_0(S_+-S_-),
\end{equation}
where the $+$ ($-$) sign refers to quantities outside, or in front of
(inside, or behind) the stagnation point. Eq.~(\ref{eq:38}) can be transformed into:
\begin{equation}
  \label{eq:39}
  \Delta G =
(s_+^{1/\gamma}-s_-^{1/\gamma})\frac{\gamma}{\gamma-1}P_s^{(\gamma-1)/\gamma}
-\frac{T_0}{\gamma-1}(\log s_+-\log s_-),
\end{equation}
where $P_s$ is the value of the pressure at the stagnation point (this value
is well-defined at this point, on the contrary to the other thermodynamics variables, since
the pressure field is continuous).
Using the first order expansions:
$s_+^{1/\gamma}-s_-^{1/\gamma}\approx(s_+-s_-)s_0^{1/\gamma-1}/\gamma$
and
$\log s_+-\log s_-\approx(s_+-s_-)/s_0$,
we are led to:
\begin{equation}
  \label{eq:40}
  \Delta G = \frac 1{\gamma-1} s_0^{1/\gamma-1}
(P_s^{(\gamma-1)/\gamma}-P_0^{(\gamma-1)/\gamma})
(s_+-s_-),
\end{equation}
where $P_0$ and $s_0$ are the pressure and the entropy at the location
of the stagnation point, in the unperturbed disk.  

In the particular case of a small perturbation:
$|P_s-P_0|\ll P_0$, the discontinuity in $G$ takes the form:
\begin{equation}
  \label{eq:41}
  \Delta G=\frac 1\gamma\frac{p_s}{\Sigma_0}\frac{s_+-s_-}{s_0},
\end{equation}
where $p_s=P_s-P_0$ is the pressure perturbation at the stagnation
point. 

In order to proceed in the evaluation of $\Delta G$, we use the
conservation of the $G$ invariant along the upstream separatrices.  At
this stage of the derivation, we note that numerical simulations
reveal that the horseshoe region in the adiabatic regime is rather asymmetric,
so that the distance of the separatrices depend on the quadrant under
consideration (rear vs. front, and upstream vs. downstream).  Denoting
with $x_u^R$ the distance of the upstream separatrix from corotation
on the rear side, we have, by evaluating $G$ respectively on that
separatrix far from the planet, by means of Eq.~(\ref{eq:19}), and at the stagnation point:
\begin{equation}
  \label{eq:42}
  G_-=G_c+2AB(x_u^R)^2,
\end{equation}
and
\begin{equation}
  \label{eq:43}
  G_-=G_c-\eta_c+\eta_-+\Phi_p^s+T_0(S_c-S_-),
\end{equation}
where $\Phi_p^s$ is the planetary potential at the stagnation point, $\eta_c$ and $S_c$ are
respectively the enthalpy and entropy at corotation in the unperturbed disk, and where we have used the
relationship $G_c=\Phi_*(r_c)-(1/2)r_c^2\Omega_p^2+\eta_c-T_0S_c$ to derive Eq.~(\ref{eq:43}).
From Eqs.~(\ref{eq:42}) and~(\ref{eq:43}) we deduce:
\begin{equation}
  \label{eq:44}
  2AB(x_u^R)^2=\eta_--\eta_c+\Phi_p^s+T_0(S_c-S_-).
\end{equation}
In a similar manner, denoting with $x_u^F$ the distance of the upstream
separatrix from corotation on the front side, we have:
\begin{equation}
  \label{eq:45}
  2AB(x_u^F)^2=\eta_+-\eta_c+\Phi_p^s+T_0(S_c-S_+).
\end{equation}
From Eq.~(\ref{eq:40}), we have:
\begin{equation}
  \label{eq:46}
  \Delta G=\frac1\gamma\left[\left(\frac{s_0}{s_\pm}\right)^{1/\gamma}\eta_\pm-\eta_c\right]
  \frac{s_+-s_-}{s_0},
\end{equation}
hence we have:
\begin{equation}
  \label{eq:47}
  2\Delta G=\frac 1\gamma(\eta_++\eta_--2\eta_c)\frac{s_+-s_-}{s_0},
\end{equation}
where we have discarded the additional term that scales with ${\cal S}^2$. Similarly,
adding Eqs.~(\ref{eq:44}) and~(\ref{eq:45}) yields:
\begin{equation}
  \label{eq:48}
  2AB[(x_u^R)^2+(x_u^F)^2]=\eta_++\eta_--2\eta_c+2\Phi_p,
\end{equation}
where we have omitted the term $T_0(2S_c-S_--S_+)$, which scales as $h^2\Omega^2x^2$,
and is hence negligible compared to the left hand side.
Using Eqs.~(\ref{eq:47}) and~(\ref{eq:48}), we obtain:
\begin{equation}
  \label{eq:49}
  \Delta G=\frac 1\gamma\left\{AB[(x_u^R)^2+(x_u^F)^2]-\Phi_p^s\right\}\frac{s_+-s_-}{s_0}
\end{equation}

Since the entropy is conserved during the advection along the upstream
separatrices, we have:
\begin{equation}
  \label{eq:50}
  \frac{s_+-s_-}{s_0}=\gamma{\cal S}\frac{x_u^F+x_u^R}{a}.
\end{equation}
The discontinuity of $G$ reads finally:
\begin{equation}
  \label{eq:51}
  \Delta G = \left\{AB[(x_u^R)^2+(x_u^F)^2]-\Phi_p^s\right\}{\cal S}\frac{x_u^F+x_u^R}{a}
\end{equation}
and the associated torque excess is, using Eq.~(\ref{eq:37}):
\begin{equation}
  \label{eq:52}
  \Gamma_{1}=-(x_u^F+x_u^R)^2\Sigma_0{\cal S}\left\{AB[(x_u^R)^2+(x_u^F)^2]-\Phi_p^s\right\},
\end{equation}
where we have used the equality $\Delta j_0^s=2Ba(x_u^R+x_u^F)$.
Since $P_s>P_0$, $\Delta G$ has same sign as ${\cal S}$, and the
torque excess scales with the negative of the entropy gradient, as
already observed in previous works \citep{bm08,pp08}.
We can render Eq.~(\ref{eq:52}) a bit more compact by defining
the averaged upstream half-width:
\begin{equation}
  \label{eq:53}
  x_s=\sqrt{\frac{(x_u^R)^2+(x_u^F)^2}{2}}
\end{equation}
Considering that the relative asymmetry between the rear and front part
of the horseshoe, albeit larger than typically observed in barotropic
situations (see paper~I), remains moderate, we have:

\begin{equation}
  \label{eq:54}
x_s\approx\frac 12(x_u^R+x_u^F),
\end{equation}
and therefore:
\begin{equation}
  \label{eq:55}
  \Gamma_{1}\approx-4x_s^2\Sigma_0{\cal S}\left(2ABx_s^2-\Phi_p^s\right),
\end{equation}
This expression displays explicitly the dependency of the torque excess on the
entropy gradient. As we shall see hereafter, it also brings in a dependency
on the potential's softening length, as both $x_s$ and $\Phi_p^s$ depend
on this quantity.

\section{Comparison to numerical simulations}
\label{sec:comp-numer-simul}
We have undertaken a number of numerical simulations in order to test
the validity of Eq.~(\ref{eq:55}), which quantifies the adiabatic torque excess. We briefly describe the code used
and the parameters adopted, then we assess the correctness of
Eq.~(\ref{eq:55}) from series of calculations in which we vary the
entropy gradient or the softening length.

\subsection{Numerical set up}
\label{sec:numerical-set-up}
We used the code FARGO\footnote{See: {\tt http://fargo.in2p3.fr}}
\citep{fargo2000,fargo2000b,bm08} in its adiabatic and isothermal
versions. Unless otherwise stated, the polar mesh extends from $r_{\rm
  min}=0.4$ to $r_{\rm max}=1.8$, and has a resolution of $1200\times
1200$, the zone boundaries being evenly spaced. The planet is located
at $r=1$, and it is held on a fixed circular orbit. Its mass ratio to
the primary is $q=3\cdot 10^{-6}$ ({\em i.e.}  the planet has one
Earth-mass, if the central object has a solar mass). The disk aspect
ratio at $r=1$ is $h=0.05$. The adiabatic index of the gas is
$\gamma=1.4$. Damping boundary conditions \citep[as described in][]{valborro06} are used at the inner and outer 
edges of the
mesh. The frame corotates with the planet, and the Coriolis force is
conservatively implemented \citep{1998A&A...338L..37K}. The surface
density is a power law of the radius: $\Sigma(r) \propto r^{-\sigma}$.

In our first set of calculations, the slope of the surface density is
set to $\sigma=+3/2$ (hence the vortensity related part of the
corotation torque cancels out), and the potential softening length
ranges from $\hat\epsilon=0.1$ to $10$, where $\hat\epsilon=\epsilon/H$ is a dimensionless value
of the softening length. The logarithm of
$\hat\epsilon$ is evenly spaced, and $21$ calculations are performed,
so that $\hat\epsilon=10^{i/10-1}$, with $0\le i\le 20$. These
calculations were performed both for an adiabatic flow (for brevity we
denote them with EA$_{i,0\le i\le 20}$) and for an isothermal flow (we
denote them with EI$_{i,0\le i\le 20}$). Since the initial temperature
field is flat, the index of the entropy gradient is $ {\cal
  S}=\sigma(\gamma-1)/\gamma=+3/7$. We therefore expect a negative
torque excess in the adiabatic case.

In our second set of calculations, the slope of the surface density
ranges from $\sigma=-4$ to $\sigma=4$, the softening length being kept
fixed at $\hat\epsilon=0.3$. We have divided the slope interval into
$41$ calculations such that $\sigma_i=-4+0.2\times i$ $(0\le i\le
40)$.  These calculations were performed both for an adiabatic flow
(we denote them with SA$_{i,0\le i\le 40}$) and for an isothermal flow
(we denote them with SI$_{i,0\le i\le 40}$). Since the initial
temperature field is flat, the index of the entropy gradient is
\begin{equation}
  \label{eq:56}
  {\cal S}=\frac{\gamma-1}{\gamma}\sigma.
\end{equation}
It therefore ranges from $-8/7$ to $+8/7$.

\subsection{A note on the half width of the horseshoe region}
\label{sec:note-half-width}
\begin{figure}
  \centering
  \plotone{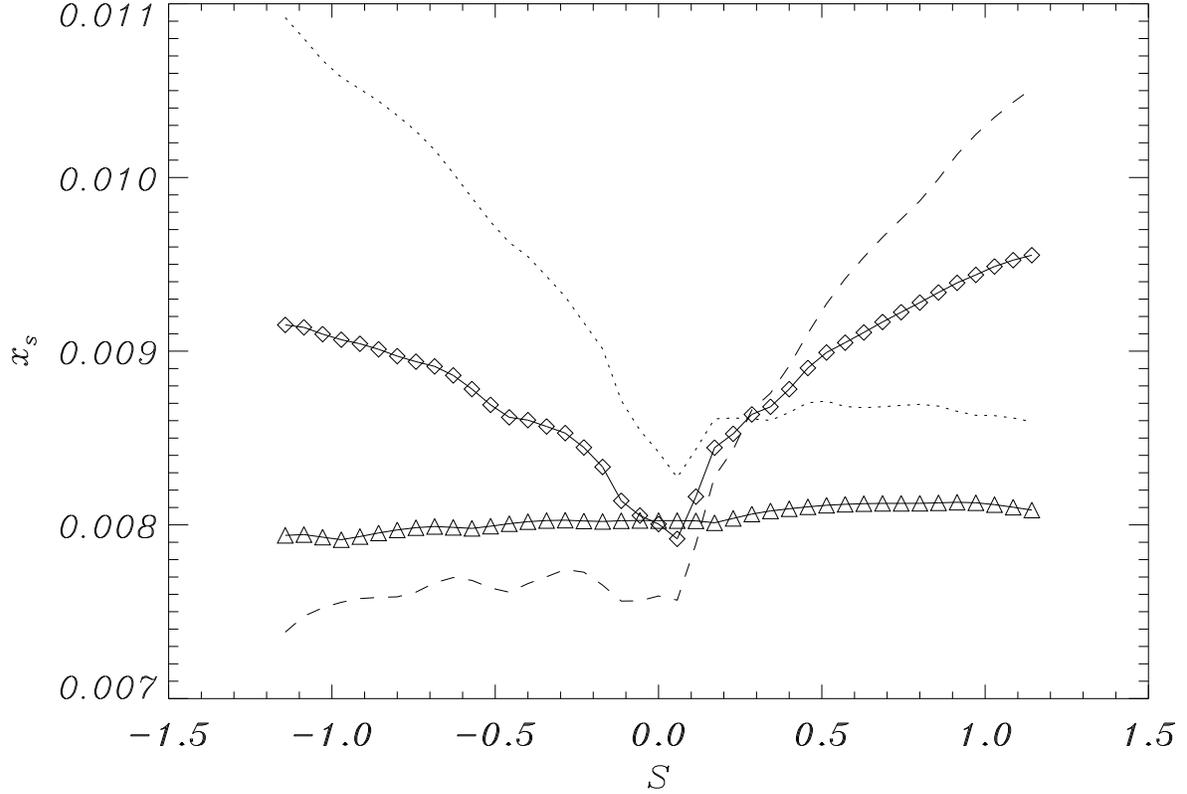}
  \caption{Half width $x_s$ of the horseshoe region in the adiabatic
    case (diamonds), as a function of the entropy gradient ${\cal S}$. 
    We use the arithmetic mean of $x_u^R$ and $x_u^F$
    as an estimate of $x_s$. The dotted line shows $x_u^F$, and the
    dashed line shows $x_u^R$. The triangles show the half width of
    the horseshoe region (divided by $\gamma^{1/4}$) obtained from
    runs with same parameters, but with an isothermal equation of
    state. The rear/front averaged upstream half width has a
    characteristic V-shape. The data presented here
    are obtained from the series of runs SA and SI.}
  \label{fig:hwhs}
\end{figure}

As underlined in section~\ref{sec:disc-at-stagn}, the horseshoe region
in adiabatic calculations is more asymmetric than in the isothermal
case. It also has a different dependence, more complex, on the disk
parameters. This can be seen in Fig.~\ref{fig:hwhs}, in which we plot the different
values of $x_s$ inferred from a streamline analysis on the runs of the series
SA$_i$. In particular,
the naive expectation that 
\begin{equation}
  \label{eq:57}
  x_s^{\rm adi}= x_s^{\rm iso}/\gamma^{1/4}
\end{equation}
is true only in the barotropic case (${\cal S}=0$). For non-vanishing
entropy gradients, the averaged upstream half width exhibits a
V-shaped dependence on ${\cal S}$, whereas the isothermal case is
insensitive to this parameter.
This more complex dependence of the width of the horseshoe region
has a consequence on the evaluation of the torque excess. A simple
estimate of the latter is:
\begin{equation}
  \label{eq:58}
  \Gamma^r_1=\Gamma^A-\frac{\Gamma^I}{\gamma},
\end{equation}
but this estimate assumes $x_{s, \rm adi}=x_{s, \rm iso}/\gamma^{1/4}$, so
that one gets rid of the vortensity related part of the horseshoe
drag, retaining only the excess. This is not exactly true, since the
adiabatic case has a horseshoe region in general larger than
expected. In order to fully eliminate the vortensity part of the
estimate, one should therefore use the expression:
\begin{equation}
  \label{eq:59}
  \Gamma^c_1=\Gamma^A-\frac{\Gamma^I}{\gamma}-\frac 34{\cal V}\Omega_p^2\Sigma_0\left(x_{s, \rm adi}^4-\frac{x_{s, \rm iso}^4}{\gamma}\right)
\end{equation}
We shall hereafter refer to the estimate given by Eq.~(\ref{eq:58}) as
the rough or global estimate (hence the $r$ superscript), and to the
estimate of Eq.~(\ref{eq:59}) as the corrected estimate (hence the $c$
superscript).

\subsection{Dependence on the softening length}
\label{sec:depend-soft-length}
We compare the torque excess, when $\hat\epsilon$ varies, as provided
by Eq.~(\ref{eq:55}) and as provided by a direct comparison of the
torque value in adiabatic and isothermal calculations.
\begin{figure}
  \centering
  \plotone{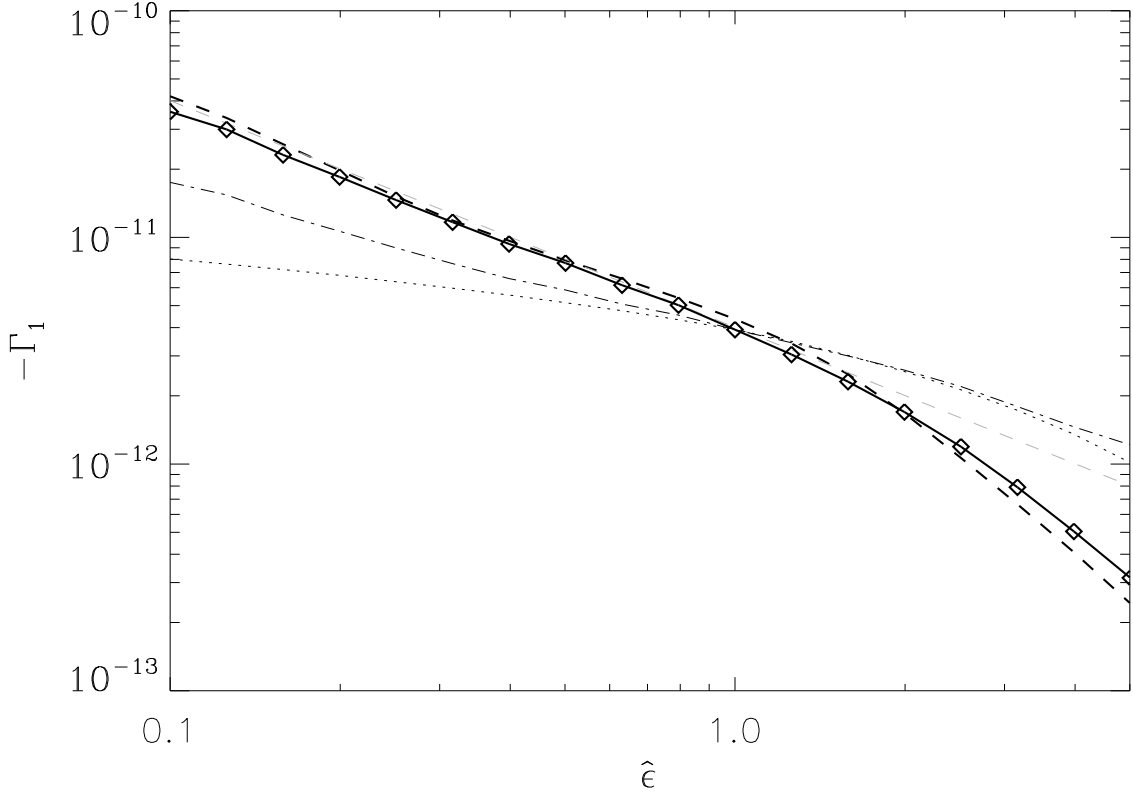}
  \caption{Torque excess as a function of the softening length. The
    thick dashed line shows the excess $\Gamma_1^c$ directly measured by comparing the runs EA$_i$
    and EI$_i$, by means of
    the Eq.~(\ref{eq:59}) --~or~(\ref{eq:58}), since the vortensity
    gradient is zero in this series of runs~--, while the thick curve
    with diamonds shows the prediction of Eq.~(\ref{eq:55}). The
    dash-dotted and dotted curves show the result of Eq.~(\ref{eq:55})
    in which one respectively adopts a constant value for $x_s$ or for
    $\Phi_p^s$ (namely the value measured for $\hat\epsilon=1$). The
    grey dashed line shows the prediction of Eq.~(\ref{eq:96}), which
    will be derived in section~\ref{sec:suit-torq-expr}.}
  \label{fig:excess-smoo}
\end{figure}
The results are presented in Fig.~\ref{fig:excess-smoo}. We see that
there is a good overall agreement between the direct measure of
$\Gamma_1^c$ and the result of Eq.~(\ref{eq:55}), as the relative
difference is at most $\sim 10$~\% over the range
$0.1\le\hat\epsilon\le 1$. The value of $x_s$, as measured on the
output at $t=40$~orbits and $\phi=\pm 1$~rad, is used in the
evaluation of Eq.~(\ref{eq:55}).  Similarly, the location of the
stagnation point is determined, and used to evaluate the planetary
potential that features in the bracket of Eq.~(\ref{eq:55}). We note
that in all the runs of this series, there is a unique stagnation
point in the planet's vicinity, hence there is no ambiguity in its
determination.

There is an approximate $\hat\epsilon^{-1}$ scaling of the torque
excess in the range $0.1\le\hat\epsilon\le 1$ \citep{phdbaruteau}. 
Fig.~\ref{fig:excess-smoo} shows that this trend results
partly from the increase of $x_s$ as $\hat\epsilon$ decreases (dotted
curve), and partly from the increase (in absolute value) of the
planetary potential as the softening length decreases (dot-dashed
line).

\subsection{Dependence on the entropy gradient}
\label{sec:depend-entr-grad}
We now evaluate how well Eq.~(\ref{eq:55}) accounts for the torque
excess in the second series of runs (SA and SI, see
section~\ref{sec:numerical-set-up}), in which we vary the entropy
gradient.
\begin{figure}
  \centering
  \plotone{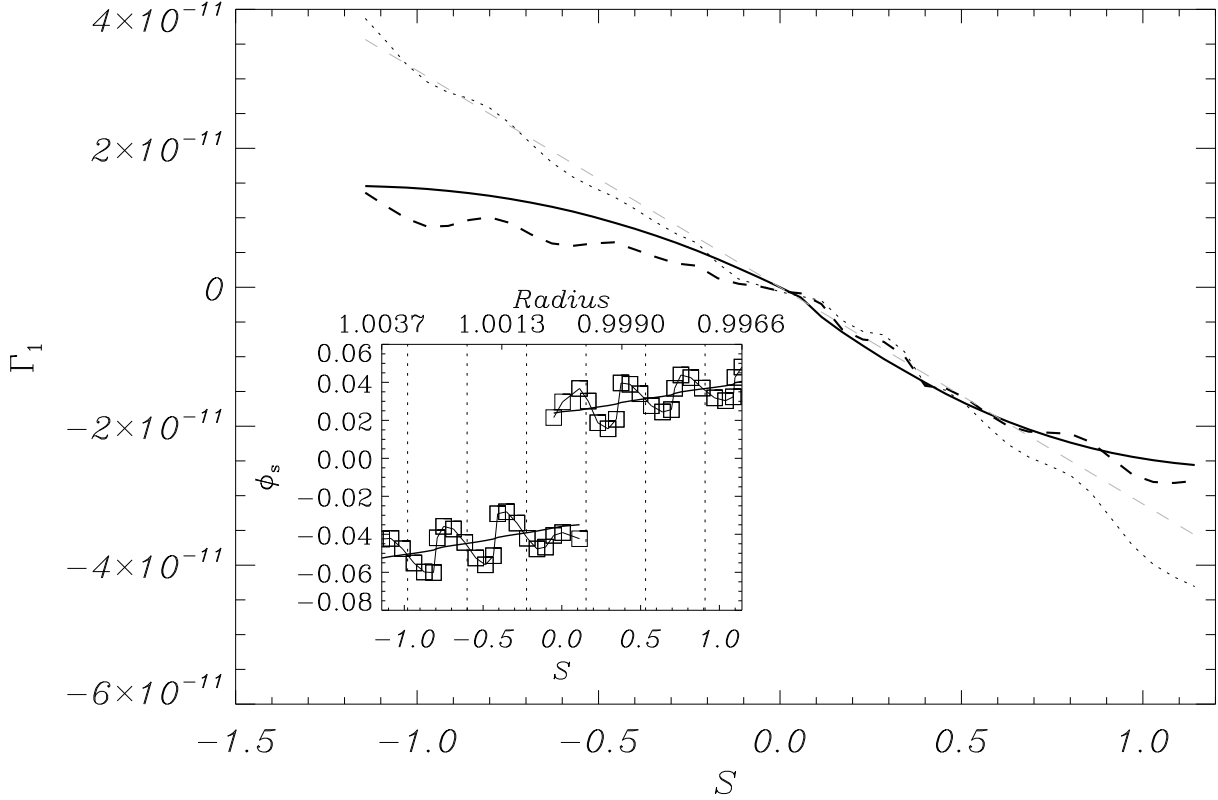}
  \caption{Torque excess as a function of the entropy gradient. The
    thick dashed line shows the corrected direct estimate $\Gamma_1^c$ given by
    Eq.~(\ref{eq:59}), while the thin dotted line shows the rough
    estimate $\Gamma_1^r$ of Eq.~(\ref{eq:58}).  The thick solid line shows the result of
    Eq.~(\ref{eq:55}), in which we use the planetary potential at the
    stagnation point and the horseshoe width, given by a prior
    streamline analysis performed at $t=40$~orbits for each
    calculation (and corrected as explained in the text).
    The inset plot shows the azimuth of the stagnation
    point, as a function of the entropy gradient (lower axis) and as a
    function of the corotation radius (upper axis). The vertical
    dotted lines show the zone boundaries.  The thick solid lines of
    the inset plot shows linear regression fits of the stagnation
    point location, aimed at getting rid of the spurious
    oscillations. These fits were used in Eq.~(\ref{eq:55}) and gave
    the thick solid curve of the main plot.  The grey dashed line
    shows the prediction of Eq.~(\ref{eq:96}), which will be derived
    in section~\ref{sec:suit-torq-expr}.}
  \label{fig:excess-entr}
\end{figure}

The results are presented in Fig.~\ref{fig:excess-entr}.  They show a
satisfactory agreement between the corrected estimate of the torque
excess (given by Eq.~[\ref{eq:59}], and represented by the thick
dashed line), and the theoretical prediction of Eq.~(\ref{eq:55}),
represented by the thick solid line. This figure also shows that the
azimuth of the stagnation point oscillates as the entropy gradient
varies. The oscillations are significantly large, and can be
unambiguously identified as a finite resolution issue, for the
following reasons:
\begin{itemize}
\item when the entropy gradient varies, the corotation radius is
  slightly shifted.  If one displays the azimuth of the stagnation
  point as a function of the corotation radius, the period of the
  oscillations is equal to the radial resolution.
\item We have carried out higher resolution calculations ($2000\times
  2000$, not shown here, as they were performed over a small number of orbits and a steady state
was not reached in the vicinity of the planet), 
  which were ran for $10$ orbits only (the
  location of the stagnation point hardly varies afterwards). They
  also yield oscillations of the azimuth of the stagnation point with
  a shorter period and smaller amplitude.
\end{itemize}
Given the large amplitude of the oscillations of the stagnation point for the resolution
that we were able to use for this series of runs, we need to smooth
the $\phi_s({\cal S})$ function. This is done simply by performing a
linear regression fit, independently for the rear and front stagnation
points, as shown in the inset plot of
Fig.~\ref{fig:excess-entr}. The fitted value
are then used in Eq.~(\ref{eq:55}), which yields the solid thick
curve.

The value of ${\cal S}=+3/7$ used in the series of runs EA and
EI corresponds to a location where the error due to the oscillations
of the azimuth of the stagnation point is small. We therefore
understand from this plot why there is a good agreement between the
measured and predicted adiabatic excess in Fig.~\ref{fig:excess-smoo}.
The fact that there is a good agreement for all values of
$\hat\epsilon$ in Fig.~\ref{fig:excess-smoo} suggests that the error on
the location of the stagnation point is essentially due to the
location of the corotation with respect to the mesh (which does not
depend on the softening length).  We recover the fact that, for
$\hat\epsilon=0.3$, $|\Gamma_1|\sim 1.5\cdot 10^{-11}$.

It is noteworthy that, on the left part of the plot of
Fig.~\ref{fig:excess-entr}, there is a sizable difference between the
rough and corrected estimates of the excess (thick dashed curve and
dotted curve). The reason for this is straightforward, since for these
values of the entropy gradient, the surface density increases steeply
outwards, hence the vortensity related corotation torque is large and
positive. For large values of $|{\cal S}|$, the adiabatic horseshoe
half width is significantly larger than $x_s^{\rm iso}/\gamma^{1/4}$
(see section~\ref{sec:note-half-width}), therefore a fair fraction of
the adiabatic excess is actually due to the boost of the vortensity
related torque.

\section{Properties of the coorbital region}
\label{sec:prop-coorb-regi}
We describe in this section the properties of the coorbital region
that differ from the barotropic case. The most salient feature is the
appearance of vorticity sheets along the downstream separatrices.  We
also derive the production of vortensity within the horseshoe
region. The latter is weak, however, and does not have any impact on
the horseshoe drag.  We finally discuss the properties of the pressure
disturbances arising from the horseshoe dynamics, and identify which
component of these is responsible for the adiabatic torque excess.

\subsection{Vorticity sheets}
\label{sec:vorticity-sheets}
\begin{figure}
  \centering
  \plotone{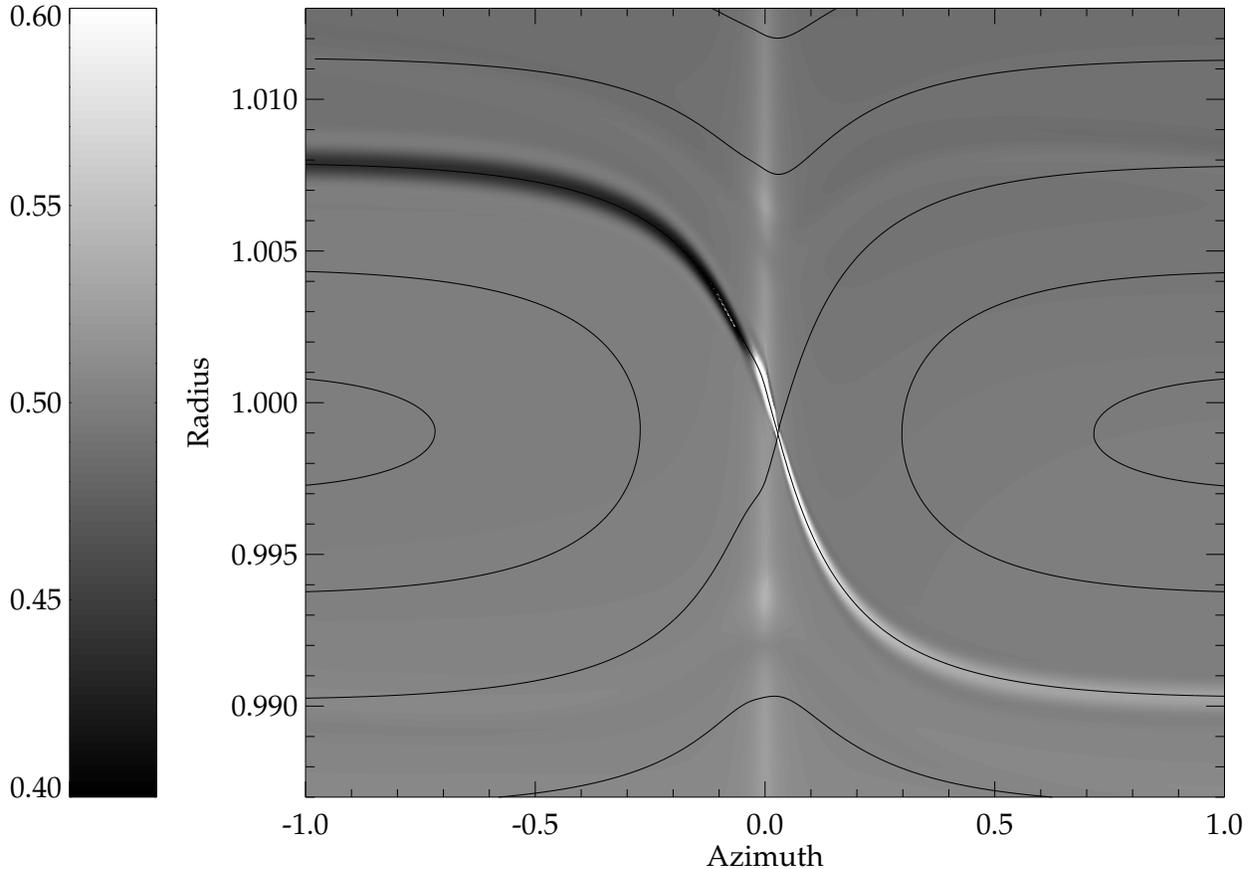}
  \caption{Vorticity field in the vicinity of an Earth-mass
    planet at $t=64$~orbits. The characteristics of the run are described in the
    text. The vorticity sheets are clearly apparent on the
    separatrices. Also, we note that the sign of the vorticity does
    not reverse at the stagnation point, but somewhere further on the
    downstream rear separatrix. The reason for this is that the
    perturbation of pressure is larger over the portion of the
    separatrix involved than at the stagnation point, as it crosses
    the wake. We also note that the vorticity is not singular on the
    upstream separatrices, which results from the fact that $G$ is
    continuous across them.}
  \label{fig:vort-field}
\end{figure}

\begin{figure}
  \centering
  \plotone{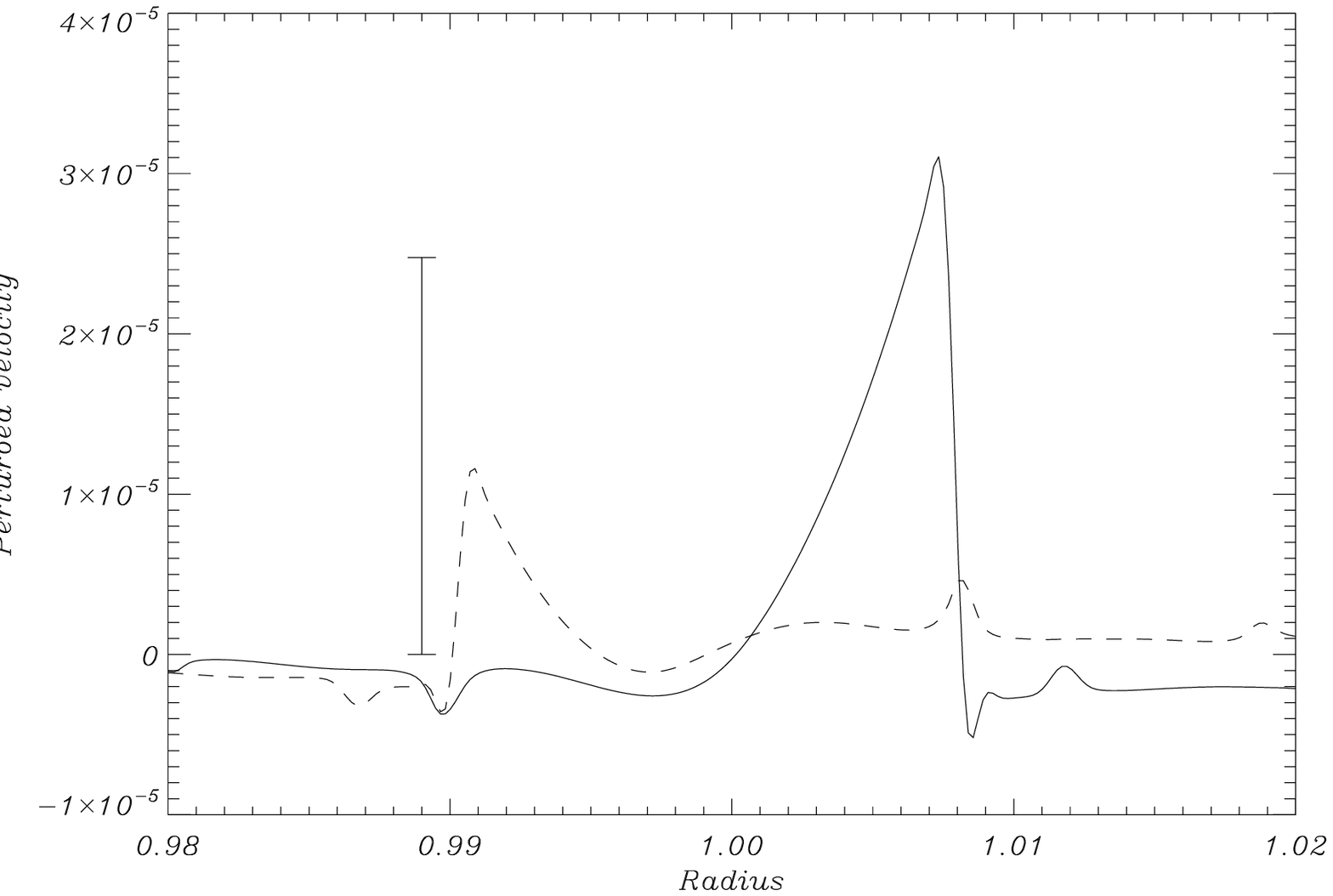}
  \caption{Perturbed azimuthal velocity at $t=64$~orbits, as a function of radius, at
    $\phi=1$~rad (dashed line) and at $\phi=-1$~rad (solid line). The
    vertical bar at the left shows the expected magnitude of the
    velocity jump, obtained from a measure of the torque excess using
    Eq.~(\ref{eq:59}), and using the result in Eq.~(\ref{eq:62}).}
  \label{fig:azim-vel-cut}
\end{figure}

The discontinuity of the $G$-invariant that we evaluated in
section~\ref{sec:disc-at-stagn} at the stagnation point also exists
along the downstream separatrices, as can be seen in
Fig.~\ref{fig:quad}. Far from the planet, one can write
the expression of the $G$ discontinuity across a downstream
separatrix as:
\begin{equation}
  \label{eq:60}
  \Delta G=\frac 1\gamma\frac{p'_s}{\Sigma_0}\frac{s_+-s_-}{s_0}+
\frac{r^2}{2}[(\Omega_+-\Omega_p)^2-(\Omega_--\Omega_p)^2],
\end{equation}
where we have performed the same series of transformations as in
Eqs.~(\ref{eq:38}) to~(\ref{eq:41}), where $p'_s$ is the perturbed
pressure at the separatrix, and where $\Omega_+$ ($\Omega_-$) is the
angular velocity on the outside (inside) of the separatrix.  If we assume
in a first approximation that the perturbation of pressure at the
separatrix, far from the planet, is much smaller than at the
stagnation point, most of the $G$-discontinuity is ensured by the jump
of azimuthal velocity, i.e. by a vorticity sheet. Denoting with
$\Delta v=r(\Omega_+-\Omega_-)$ the magnitude of the vorticity sheet
(i.e. the singular part of the flow's vorticity reads: $\omega=\Delta
v\delta(x-x_s)$, where $\delta$ is Dirac's delta function), and
assuming that the jump is sufficiently small so that we can write:
\begin{equation}
  \label{eq:61}
  \Omega_++\Omega_--2\Omega_p=\frac{4Ax}{a},
\end{equation}
we are led to:
\begin{equation}
  \label{eq:62}
  \Delta v=\frac{\Delta G}{2Ax}=
\left\{
\begin{array}{ll}\displaystyle\frac{\Gamma_1}{4a|A|x_s^2\Sigma_0}&\mbox{~at the rear separatrix}\\
\displaystyle-\frac{\Gamma_1}{4a|A|x_s^2\Sigma_0}&\mbox{~at the front separatrix}
\end{array}\right.
\end{equation}
Therefore, if there is a positive torque excess (hence ${\cal S}<0$),
then there is a negative (positive) vorticity sheet in front of the
planet (at the rear of the planet), and vice-versa. In order to
illustrate this, we show in Fig.~\ref{fig:vort-field} the vorticity
field in a run with ${\cal S}=1/3$ and similar characteristics as those
of section~\ref{sec:numerical-set-up}, except that the resolution was
set to a $2000$ zones in azimuth and $8000$ zones in radius.
We measure an adiabatic torque excess $\Gamma_1=-1.14\cdot
10^{-11}$~$M_*\Omega_p^2r_p^2$, by performing a run with similar
characteristics and an isothermal equation of state. This allows us to
predict the jump of angular velocity at the downstream separatrices by
the use of Eq.~(\ref{eq:62}).  The results are depicted in Fig.~\ref{fig:azim-vel-cut}, which
shows the profile of perturbed azimuthal velocity in the rear and
front of the planet, as well as the prediction of
Eq.~(\ref{eq:62}).
We note that the jumps are not symmetric, although there is a good
agreement between the theoretical prediction of the velocity jump on
the one hand, and the average value of the rear and front velocity
jumps on the other hand. One reason for that is that the perturbed
pressure at the separatrices is not negligible compared to the
perturbed pressure at the stagnation point. Another one is the
asymmetry of the horseshoe region: both separatrices do not have
same value of $|(\Omega_++\Omega_-)/2-\Omega_p|$.

We note that for the run presented here, for which ${\cal S}>0$, we
have a negative torque excess, hence a negative perturbation of
surface density in front of the planet, and a positive one behind the
planet \citep{bm08}.  The fact that we have a positive perturbation of
azimuthal velocity in front of the planet (dashed curve of
Fig.~\ref{fig:azim-vel-cut}) is compatible with an expansion of the
material and consequently a decrease of surface density, because this
perturbed velocity acts along with the Keplerian shear. Reciprocally, we
have also a positive perturbation of velocity behind the planet,
therefore opposed to the Keplerian shear, and compatible with a
compression of material, hence an increase of the surface density.

Finally, we mention that the vortensity is ill-defined at the separatrices.
Whereas the vorticity is definable, the integral of its singular part yielding
the azimuthal velocity jump, the vortensity would appear, on the separatrix,
as the ratio of a singularity (that of the vorticity) by a non-continuous function
(owing to the contact discontinuity, the surface density undergoes a jump at
the separatrix), which is mathematically undefined. On the contrary, in the
locally isothermal case (see paper~I), the surface density field is continuous
and the vortensity is defined everywhere.

\subsection{Production of vortensity}
\label{sec:prod-vort}
On the contrary to a barotropic situation, for which the vortensity is
necessarily conserved along the streamlines \citep{lovelace99}, some
thermodynamical driving of the vortensity may occur during the
horseshoe U-turns. We also note that far from these U-turns, the
vortensity is conserved again, since all quantities depend only on
$r$, hence the pressure and density gradients are both radial, hence
aligned.

Using Eq.~(\ref{eq:23}) and~(\ref{eq:27}), we are left with:
\begin{equation}
  \label{eq:63}
  \left.\frac{\Sigma}{\omega}\right|_d
  =
  \left.\frac{\Sigma}{\omega}\right|_u
  \left(1-\delta T\frac{\partial_rS_0}{4ABx}\right).
\end{equation}
We comment that we have used the relationship
$\partial_SG=-\partial_rS_0/(4ABx)$ to obtain Eq.~(\ref{eq:63}). The
bracket of this equation corresponds indeed to material downstream of a
horseshoe U-turn. During a U-turn, $\partial_GS$ is conserved, while
$x$, to lowest order, reverses its sign, and Eq.~(\ref{eq:23}) has
been written for the unperturbed disk, that is to say upstream of
horseshoe U-turns.  The perturbation of temperature, within the
horseshoe region, can be easily worked out assuming the conservation
of entropy and an unperturbed pressure. This yields \citep{bm08}:
\begin{equation}
  \label{eq:64}
  \delta T=-2T_0{\cal S}\frac{x}{r_p}.
\end{equation}
Using Eqs.~(\ref{eq:63}) and~(\ref{eq:64}) we get the production of vortensity
during a U-turn:
\begin{equation}
  \label{eq:65}
  \delta\left(\frac\Sigma\omega\right)=\left.\frac{\Sigma}{\omega}\right|_d
-\left.\frac{\Sigma}{\omega}\right|_u\approx \frac{\gamma T_0{\cal S}^2}{2(\gamma-1)ABr_p^2}\cdot \frac{\Sigma_0}{2B}.
\end{equation}
\begin{figure}
  \centering
  \plotone{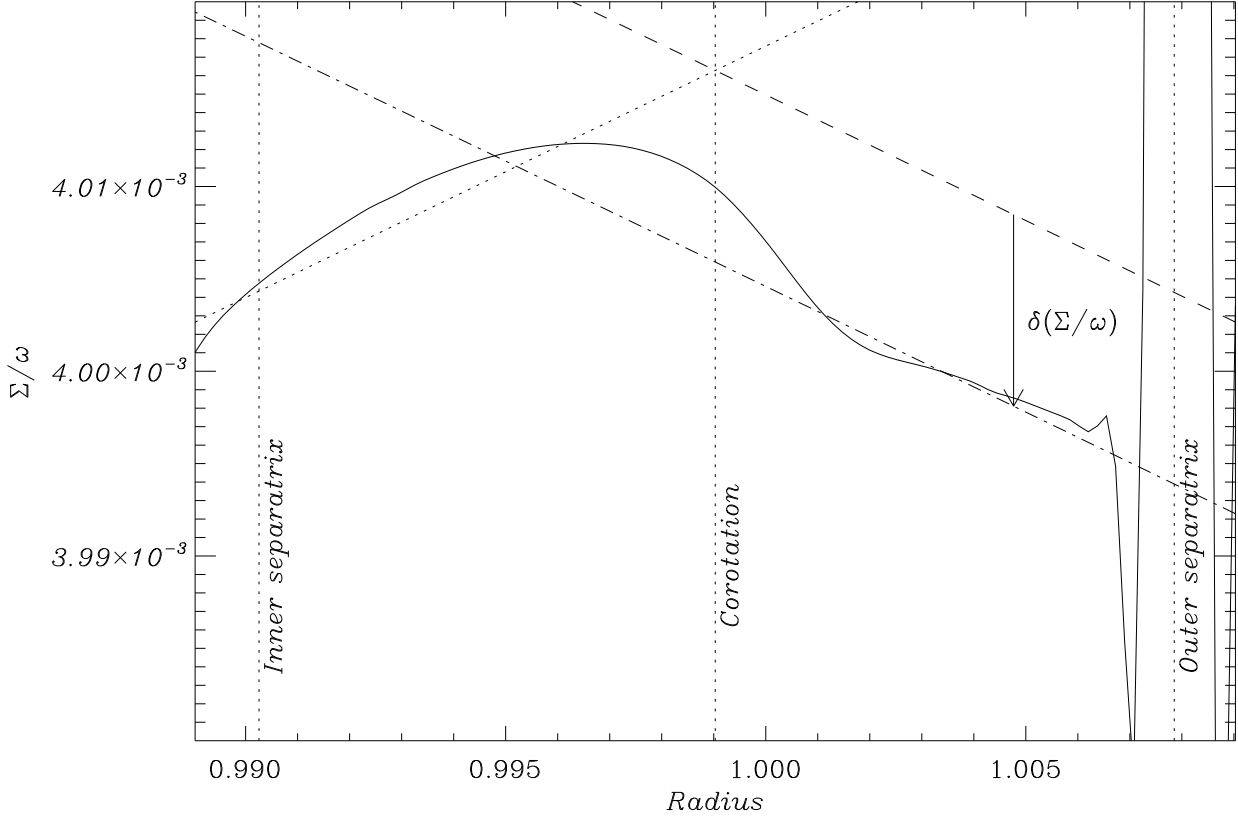}
  \caption{Radial profile of $\Sigma/\omega$ at $t=48$~orbits and
    $\phi=-1$~rad (solid line). The tilted dotted line shows the
    initial, unperturbed profile, while the tilted dashed line shows
    this profile symmetrized with respect to the corotation radius. It
    is therefore this profile that one would except if vortensity were
    conserved. If one takes into account the production of vortensity
    given by Eq.~(\ref{eq:65}), one obtains the dot-dashed line, which
    is in correct agreement with the observed profile.}
  \label{fig:cut-vortens-prod}
\end{figure}
We illustrate this in Fig.~\ref{fig:cut-vortens-prod}, which shows
that there is a satisfactory agreement between the vortensity on the
downstream side (here, outside, at the rear of the planet) and the
vortensity predicted by Eq.~(\ref{eq:65}).  We also note in this
figure how the vortensity near corotation differs from the expected
value, both on the inside and on the outside. On the outside, this is
expected because the material very close to corotation did not have
sufficient time to perform a horseshoe U-turn. On the inside, for
$0.996\le r\le 0.999$, libration on tadpole streamlines brings
vortensity from the outside, thereby yielding a perturbation of
vortensity in the upstream region. More generally, the fact that
streamlines are not exactly circular alters the initial profile of
vortensity in this region.  We also note the spike at the outer
separatrix, resulting from the vorticity sheet presented in
section~\ref{sec:vorticity-sheets}, where we also pointed out that,
formally, the vortensity is ill-defined on the downstream
separatrices.

We can make the following observations regarding Eq.~(\ref{eq:65}):
\begin{itemize}
\item There is always a positive production of vortensity during the
U-turns (hence a decrease of $\Sigma/\omega$), regardless of the
sign of ${\cal S}$, and regardless of the location (at the rear or in
front of the planet).
\item The production of vortensity is independent of the planet mass,
  and it occurs uniformly all over the downstream sides of the
  horseshoe region. While wider horseshoe U-turns bring the material
  closer to the planet, where the driving of vortensity is more
  efficient because the gradients of pressure and density are larger,
  they are performed faster than the more narrow U-turns, which occur
  further away from the planet. Eq.~(\ref{eq:65}) shows that both
  effects cancel out, resulting in a flat profile of vortensity
  perturbation.
\item If we translate the production of vortensity into a perturbation
  of density, assuming an unperturbed velocity field, then this
  perturbation has same sign on both sides of the planet. This is
  compatible with the fact that the bulk term of the horseshoe drag,
  i.e. the first term of Eq.~(\ref{eq:37}), is the same as in a
  barotropic situation. The weak thermodynamical driving of vortensity
  observed within the horseshoe region has therefore no impact on the
  torque.
\item The increase of vortensity after the first horseshoe U-turn is
  followed by a decrease of same magnitude at the following
  U-turn (in the idealized, inviscid situation considered here). 
  This can be seen from Eq.~(\ref{eq:27}), in which we revert
  the $d$ and $u$ indexes. The vortensity of a given fluid element
  therefore oscillates between its initial upstream value and its
  subsequent downstream value, and it does not undergo any drift on
  the long term.
\end{itemize}

\subsection{Pressure disturbances}
\label{sec:press-dist}
In this section we wish to determine the shape and amplitude of the
pressure disturbances which arise in the coorbital region as the
result of the horseshoe dynamics. In paper~I, we have seen that the
perturbation of vortensity was entirely accounted for by evanescent
waves, so that the density or pressure response in the coorbital
region was smoothed over a length-scale $H$. Here, we expect that the
perturbation will be split into a localized part, which ends at the
contact discontinuity at the separatrix, and a smooth part,
corresponding to the launch of evanescent waves.

Far from the planet, we assume purely circular motion, hence we can
describe the disk's perturbed state with the perturbed azimuthal
velocity $\delta v_\phi(x)$, the perturbed surface density
$\delta\Sigma(x)$, and the perturbed pressure $\delta P(x)$. We search
the relationship between this set of perturbed functions, and the
perturbations of vortensity $\delta w(x)$ and entropy $\delta
s(x)$. We note that $x$ has an arbitrary sign  for
the rest of this section. If we consider a radial
profile in front of the planet, where the downstream separatrix lies
inside of corotation, then $x_s$ is meant to be a negative quantity.

Using Eq.~(\ref{eq:12}), the disk's rotational equilibrium yields:
\begin{equation}
  \label{eq:66}
  -2\Omega_à\delta v_\phi+\frac{\partial_x\delta P}{\Sigma_0}
+\frac{\partial_xp_0}{\Sigma_0}\frac{\delta\Sigma}{\Sigma_0}=0
\end{equation}
Anticipating on what follows, we note that while the radial derivative
of an unperturbed quantity $\xi_0$ is of order $\xi_0/r$, the radial
derivative of a perturbed quantity $\delta\xi$ is of order
$\delta\xi/H$. The second term of Eq.~(\ref{eq:66}) is therefore $r/H$
times larger than the third one. This equation can therefore be simplified
as:
\begin{equation}
  \label{eq:67}
  -2\Omega_0\delta v_\phi+\frac{\partial_x\delta P}{\Sigma_0}=0
\end{equation}
Denoting $l=\Sigma/\omega$ the inverse of the
vortensity, we can write:
\begin{equation}
  \label{eq:68}
  \delta l=\frac{\delta\Sigma}{\omega_0}-\frac{\Sigma_0}{\omega_0^2}\partial_x\delta v_\phi,
\end{equation}
while the perturbation of entropy reads:
\begin{equation}
  \label{eq:69}
  \delta s=\frac{\delta P}{\Sigma_0^\gamma}-\frac{\gamma s_0}{\Sigma_0}\delta\Sigma.
\end{equation}
Deriving Eq.~(\ref{eq:67}) with respect to $x$, keeping only the
derivatives of rapidly varying terms, i.e. the perturbations, and
using Eqs.~(\ref{eq:68}) and~(\ref{eq:69}) so as to keep only the
variable $\delta P$, we obtain the following differential equation:
\begin{equation}
  \label{eq:70}
  \delta P-\frac{c_s^2}{\kappa^2}\partial_{x^2}^2\delta P = c_s^2\Sigma_0
\frac{\delta  u}{u_0},
\end{equation}
where $\kappa=(2\Omega_0\omega_0)^{1/2}$ is the epicyclic frequency,
$u=s^{1/\gamma}l$, and where we recall that $c_s$ is the
adiabatic speed of sound (see section~\ref{sec:notation}).  The
general solution of Eq.~(\ref{eq:70}) is the convolution product of
its right hand side by the Green's kernel $K(x)$, which is the
solution of:
\begin{equation}
  \label{eq:71}
  \delta P-\frac{c_s^2}{\kappa^2}\partial_{x^2}^2\delta P = \delta(x),
\end{equation}
and whose expression is:
\begin{equation}
  \label{eq:72}
  K(x) = (2\gamma^{1/2}H)^{-1}e^{-|x|/(\gamma^{1/2}H)},
\end{equation}
where we have specialized to the Keplerian case, for which $H\equiv
c_s^{\rm iso}/\Omega=c_s/(\gamma^{1/2}\kappa)$.  This kernel
represents the pressure response to a singular perturbation of $u$ at
$x=0$, of weight $\int u(x)dx=u_0/(c_s^2\Sigma_0)$. We note that
$K(x)$ has a unitary weight:
\begin{equation}
  \label{eq:73}
  \int_{-\infty}^{+\infty}K(x)dx = 1.
\end{equation}
Once the pressure response is known, one can
infer the density response from Eq.~(\ref{eq:69}):
\begin{equation}
  \label{eq:74}
  \delta\Sigma=\frac {\Sigma_0}{\gamma}\left(\frac{\delta P}{P_0}-\frac{\delta s}{s_0}\right).
\end{equation}
Denoting $\delta\Sigma_d=\delta P/c_s^2$ and $\delta\Sigma_l=-\Sigma_0\delta s/(\gamma s_0)$,
we have:
\begin{equation}
  \label{eq:75}
  \delta\Sigma=\delta\Sigma_d+\delta\Sigma_l.
\end{equation}
In Eq.~(\ref{eq:75}), the subscript $d$ stands for {\em diffuse}, since this component of the
surface density scales with the perturbed pressure, which is smoothed over a length-scale $H$
by the action of evanescent waves, as it involves the convolution of an arbitrary function by the
kernel $K$ defined at Eq.~(\ref{eq:72}). Similarly, the subscript $l$ stands for {\em localized}, since
this component of the surface density scales with the perturbation of entropy, which is advected by
the flow, and remains localized within the horseshoe region.

Since $\delta P$ is the convolution product of $K$ by $c_s^2\Sigma_0\delta u/u$, we have:
\begin{equation}
  \label{eq:76}
  \delta\Sigma_d=K\ast\Sigma_0 \frac{\delta u}{u_0},
\end{equation}
where $\ast$ denotes the convolution product. Similarly, we have:
\begin{equation}
  \label{eq:77}
  \delta\Sigma_l=-\frac{\Sigma_0}{\gamma}\frac{\delta s}{s_0}.
\end{equation}
The expression of the entropy perturbation is straightforward, as the latter is simply advected by
the horseshoe dynamics. Since $s\propto r^{\gamma{\cal S}}$, we have $\delta s/s_0=-2\gamma{\cal S}x/r_p$,
hence \citep{bm08}:
\begin{equation}
  \label{eq:78}
  \delta\Sigma_l=2\Sigma_0{\cal S}x/r_p.
\end{equation}
The expression of the perturbation of $u$ involves three components:
the singular contribution of the vorticity sheet at the separatrix,
the advection of the vortensity and entropy by the horseshoe dynamics,
and the production of vortensity given by Eq.~(\ref{eq:65}).  We have
mentioned in section~\ref{sec:vorticity-sheets} that the vortensity is
ill-defined on the downstream separatrices, since the surface density
is discontinuous there. Nevertheless, the jumps of surface density are
of order $x_s/a$, hence to lowest order the vortensity can be
estimated to be $\Delta v\delta(x-x_s)/\Sigma_0$ at the
separatrices. We can therefore estimate the perturbation in $u$:
\begin{equation}
  \label{eq:79}
  \frac{\delta u}{u_0}=-\frac{\Delta v\delta(x-x_s)}{\omega_0}-\frac{2x}{r_p}({\cal S}+{\cal V})
+\frac{\delta l_{\rm prod}}{l_0},
\end{equation}
where we have used the fact that $u\propto r^{{\cal S}+{\cal V}}$, and where, using Eq.~(\ref{eq:65}):
\begin{equation}
  \label{eq:80}
  \frac{\delta l_{\rm prod}}{l_0}=\frac{\gamma T_0{\cal S}^2}{2(\gamma-1)ABr_p^2}.
\end{equation}
For the sake of brevity, in what follows, we will keep the notation
$\delta l_{\rm prod}$, remembering that it corresponds to a uniform
production of vortensity that scales with ${\cal S}^2$, and which has
same sign on both sides of the planet. The perturbation of density finally reads:
\begin{equation}
  \label{eq:81}
  \delta\Sigma=K\ast\left[-l_0\Delta v\delta(x-x_s)-\frac{2x}{r_p}\Sigma_0({\cal S}+{\cal V})
+\frac{\Sigma_0\delta l_{\rm prod}}{l_0}\right]+\frac{2x}{r_p}\Sigma_0{\cal S}.
\end{equation}
Eq.~(\ref{eq:81}) features two kinds of terms:
\begin{itemize} 
\item bulk terms (second to fourth term), which are defined for $0\leq x\leq x_s$, prior
to a possible convolution by $K$;
\item an edge term (the first one), which is defined at $x=x_s$, prior to the convolution by $K$.
\end{itemize} 

\begin{figure}
  \centering
  \plotone{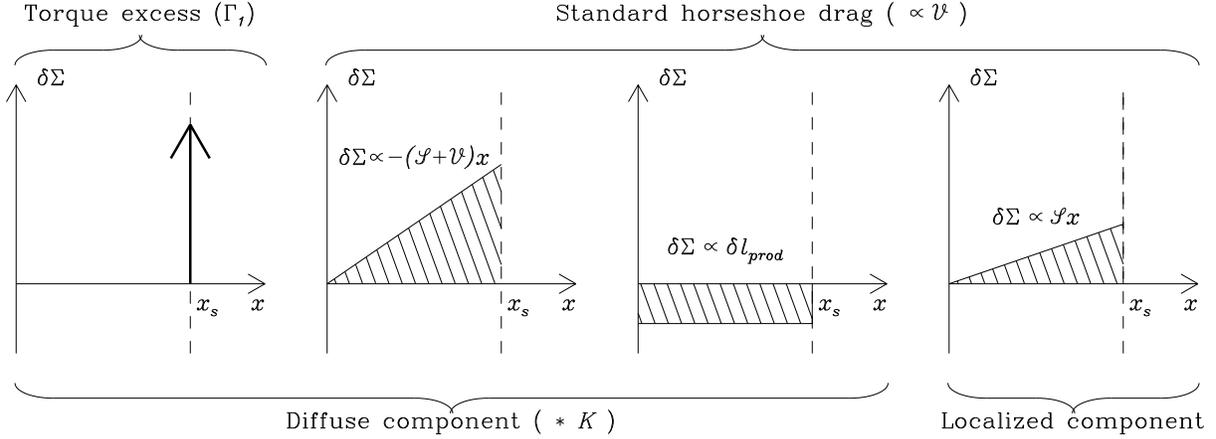}
  \caption{Schematic representation of the different terms of
    Eq.~(\ref{eq:81}), in their order of appearance in the right hand
    side.  The vertical arrow on the left plot represents the
    singularity at the separatrix. The first three terms represent the
    pressure supported density response, which is obtained by a
    convolution of the response depicted by the evanescent wave kernel
    of Eq.~(\ref{eq:72}). The last term remains confined to the
    horseshoe region and is associated to the entropy wave. The terms
    are here depicted for $x>0$, i.e. behind the planet. In front of the
    planet, they would have an opposite sign, except the third term,
    corresponding to the production of vortensity, which would remain
    negative and has therefore no impact on the torque.}
  \label{fig:schema-terms}
\end{figure}

We better understand why the bulk of the horseshoe drag has
the same expression as in the barotropic case. The convolution by $K$,
which describes the spread of disturbances by evanescent waves, does
not change the linear mass ($\int_x\delta\Sigma(x)dx$) of the
perturbation, since $K$ has a unit weight. Therefore, the torque
exerted on the planet by the diffuse and localized stripes of
perturbed density is the same as if all the disturbances remained
localized, i.e. as if we omitted the convolution product in
Eq.~(\ref{eq:81}). If we discard the singularity, which corresponds
to the edge term (and, as we shall see, to the adiabatic torque excess), 
and the production of vortensity, which has same
sign on both sides of the planet and does not affect the torque, we
are left with the middle term of the bracket of Eq.~(\ref{eq:81}),
and the trailing, localized term, which simplify as:
\begin{equation}
  \label{eq:82}
  \delta \Sigma=-\frac{2x}{r_p}\Sigma_0{\cal V},
\end{equation}
exactly as in the barotropic case, prior to the convolution by the
Green's kernel (see paper~I). These properties are schematically depicted in
Fig.~\ref{fig:schema-terms}.

This first term of Eq.~(\ref{eq:81}) corresponds to a single
evanescent wave excited at the separatrix. It is the perturbed density
associated with this wave that exerts the adiabatic torque excess,
since it is the only component that scales with the entropy gradient,
owing to the simplification mentioned above. This is in agreement
with Eq.~(\ref{eq:37}), which shows that the torque excess comes from an edge effect rather
than a bulk effect, since it relies on a minute difference between the limit of the integration domain
between the rear and front sides.  We
can check from Eq.~(\ref{eq:62}) that this edge term has correct sign. Since the
perturbed density associated to this wave has a sign opposite of
$\Delta v$, the perturbed density has same sign as $\Gamma_1$ in front
of the planet, and an opposite sign behind the planet, as expected.

A consequence of these properties is that, contrary to early
expectations, it is not the localized component of the surface
density that exerts the torque excess. The latter does exert a torque
on the planet, which can be estimated either by means of a partial
horseshoe drag calculations \citep{bm08}, or by direct summation
\citep{pp08}. It is found to have same sign as the excess, and same
order of magnitude in usual setups. Nevertheless, this partial torque
has to be added to the torque exerted by the diffuse component which
scales as ${\cal S}+{\cal V}$, and the result is a corotation torque
that scales only with the vortensity gradient. This can be expected on
general grounds. The splitting of the perturbation into a localized
component (an entropy wave) and pressure supported waves is very
similar to the procedure used by numericists who use characteristic
tracing to predict Riemann states, except that here the pressure
supported waves are evanescent instead of propagative. The entropy
perturbation regulates how much of the perturbation goes into the
entropy wave, but it is the vortensity perturbation that dictates the
total amount of mass perturbation, and it is therefore logical that the
bulk of the corotation torque scales with the vortensity gradient.

\begin{figure}
  \centering
  \plotone{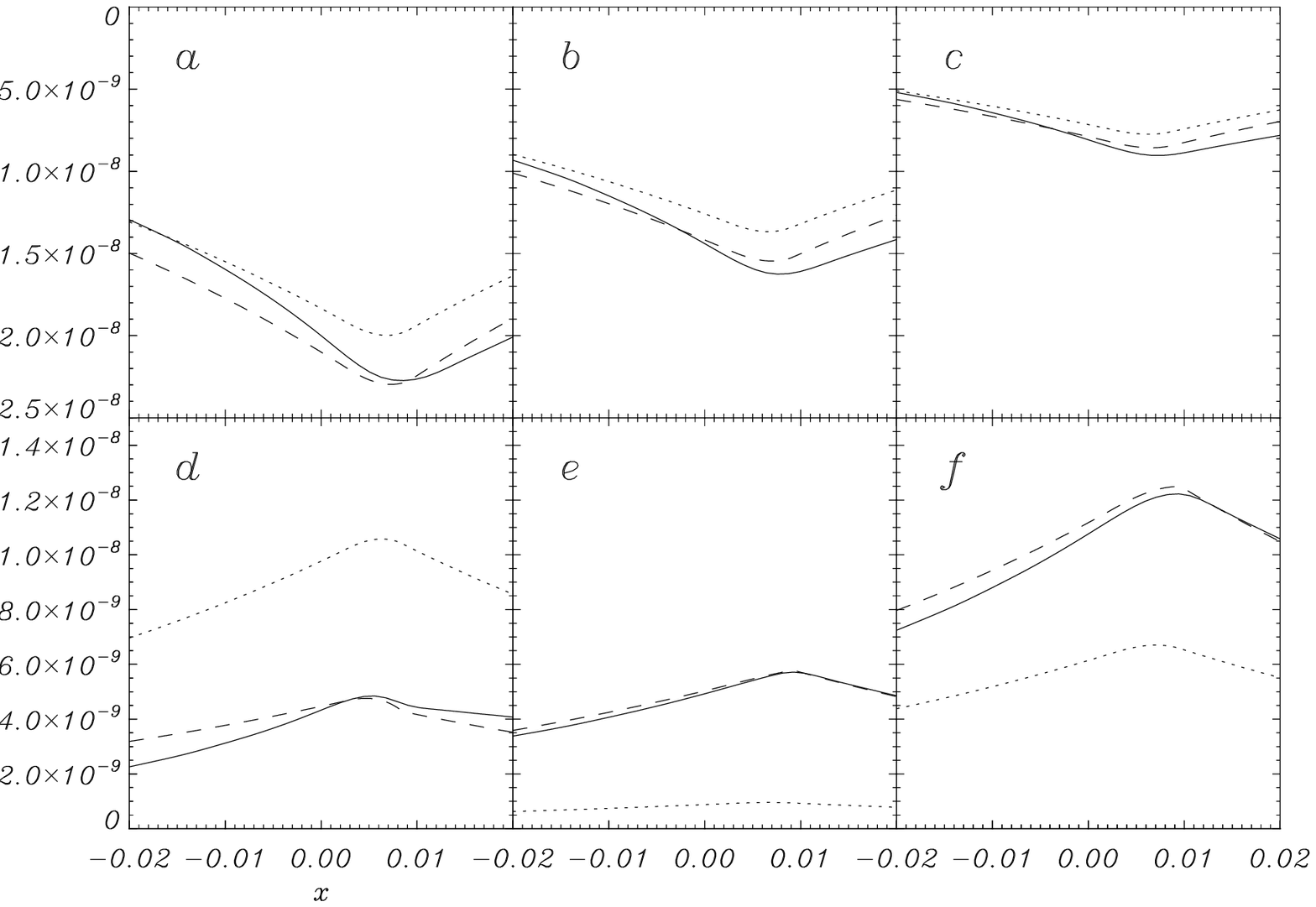}
  \caption{Profile of pressure perturbation for the calculations
    EA$_0$ $(a)$, EA$_8$ $(b)$, EA$_{16}$ $(c)$, EA$_{24}$ $(d)$,
    EA$_{32}$ $(e)$ and EA$_{40}$ $(f)$. The solid line shows the
    profile observed in numerical simulations, evaluated using
    Eq.~(\ref{eq:83}), and the dashed line shows the result of
    Eq.~(\ref{eq:109}). The dotted line shows the partial result of
    Eq.~(\ref{eq:109}), in which we omit the first term, which
    corresponds to the isolated evanescent wave at the separatrix
    corresponding to the torque excess.  The results of the plot $d$
    have been multiplied by $-3$, in order to make it more
    legible. This case corresponds to an approximate cancellation of
    the pressure term stemming from the ${\cal S}+{\cal V}$ term and
    of the single evanescent wave excited at the separatrix. The case
    of the plot $e$ is also of interest: for this calculation, ${\cal
      V}+{\cal S}\approx 0$, as can be seen from the dotted line,
    hence the pressure perturbation almost exclusively consists of the
    single evanescent wave excited at the separatrix, which is
    responsible for the torque excess.}
  \label{fig:prof-pert-pres-EA}
\end{figure}

Owing to the simplicity of the functions involved in the convolution
product, it is possible to write explicitly the expression of the
pressure perturbation. This is done in
appendix~\ref{sec:expr-press-pert}. We show in
Fig.~\ref{fig:prof-pert-pres-EA} that the resulting expression, given
by Eq.~(\ref{eq:109}), satisfactorily reproduces the perturbations of
pressure observed in numerical simulations. Namely, we transform the
profile of perturbed pressure from a numerical simulation as follows,
in order to filter out the term arising from the production of vortensity,
which is poorly reproduced by a rectangular function, as it has been
shown in Fig.~\ref{fig:cut-vortens-prod}:
\begin{equation}
  \label{eq:83}
  \delta P_{\rm sym}(x) = \frac12[\delta P(\phi=2\pi-1,x)-\delta P(\phi=1,-x)]
\end{equation}
We then evaluate the theoretical expression of the perturbed pressure
given by Eq.~(\ref{eq:109}), in which we use for $\Gamma_1$ the
estimate provided by Eq.~(\ref{eq:59}) (corresponding to the thick
dashed curve of Fig.~\ref{fig:excess-entr}), and for $x_s$ the
arithmetic mean of $x^d_F$ and $x^d_R$, determined by a streamline
analysis. Owing to the symmetrization performed in Eq.~(\ref{eq:83}),
we discard the term in $\delta l_{\rm prod}$ in Eq.~(\ref{eq:109}).
The correct agreement between the observed and predicted pressure
profile indirectly confirms that it is indeed the single evanescent
wave excited at the separatrix that is responsible for the torque
excess. We also recover a feature noticeable in
Fig.~\ref{fig:excess-entr}. While the plots~$a$ and~$f$ of
Fig.~\ref{fig:prof-pert-pres-EA} correspond to the same absolute value
of the entropy gradient (namely $8/7$), the single evanescent wave has
a much larger amplitude in the case ${\cal S}>0$ (i.e. for the
plot~$f$), almost of a factor of two, as can be seen from the distance
between the dotted and dashed curves. This is compatible with the fact
that, in Fig.~\ref{fig:excess-entr}, we found that the corrected torque
excess, in absolute value, is almost a factor of two larger at ${\cal S}=8/7$
than at ${\cal S}=-8/7$.

\section{Interpretation: an intrinsic asymmetry}
\label{sec:interpr-an-intr}
\begin{figure}
  \centering
  \plotone{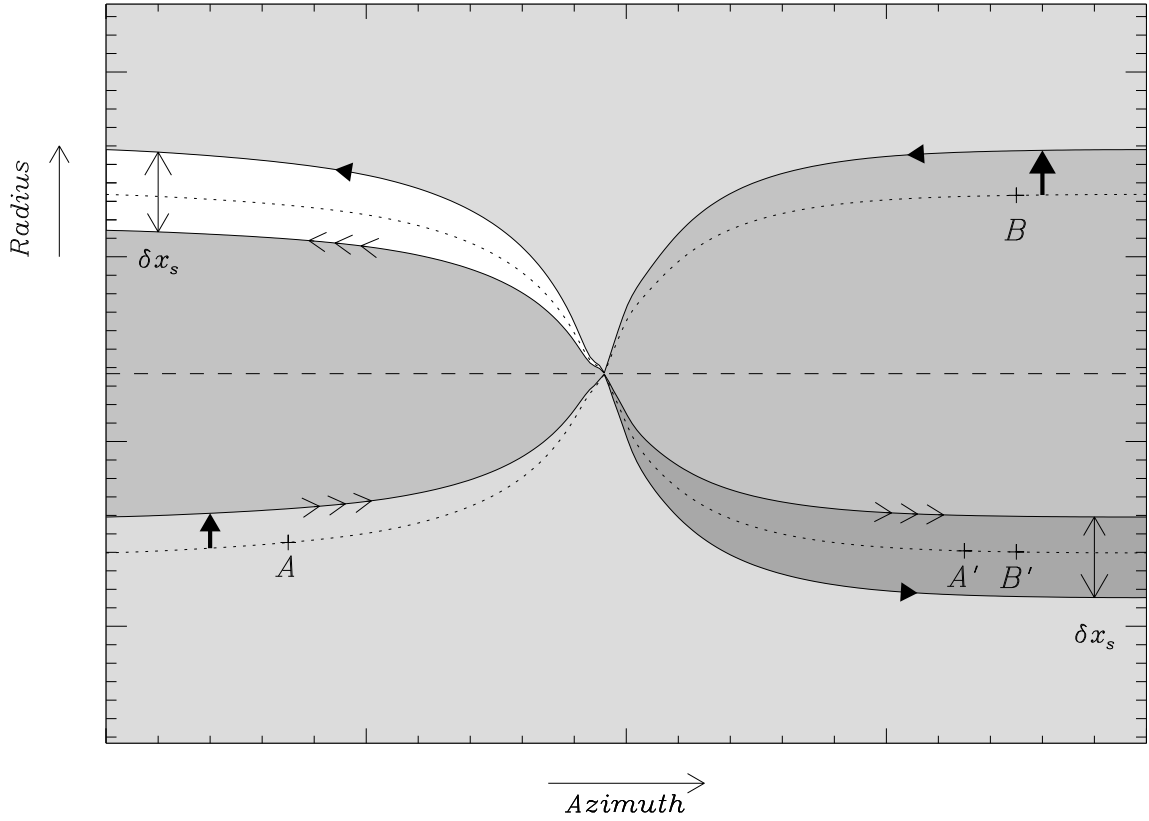}
  \caption{Schematic representation of an asymmetric horseshoe region.
    The dotted lines represent the symmetric separatrices of a
    barotropic situation. With respect to this case, the rear
    separatrix is shifted towards corotation, while the front
    separatrix is shifted away from corotation. The width
    $\delta x_s$ of the stripes has been exaggerated to improve
    legibility.}
  \label{fig:scheme-asym}
\end{figure}

We have seen in paper~I how the excitation of evanescent pressure
waves alters the width of the horseshoe region, in a different manner
in front of the planet and behind the planet, so that it renders the
horseshoe region asymmetric. This asymmetry, however, was found to
have virtually no impact on the horseshoe drag, while the separatrices
all shared the same value of the Bernoulli invariant.  In the adiabatic
case considered here, the evanescent waves launched downstream of the
U-turns (see section \ref{sec:press-dist}) also alter the horseshoe
width and render the horseshoe region asymmetric, but there is another
source of asymmetry that preexists the launch of evanescent waves. Let
us assume for a moment that the disk is composed of non-interacting
test particles, so that there are no evanescent waves excited by the
horseshoe motion, but let us also assume that the separatrices of the
horseshoe region are given by the $G_+$ and $G_-$ values, as in an
adiabatic disk (the expansion of $G$ to lowest order in $x$, given by
Eq.~(\ref{eq:19}), is independent of the disk's thermodynamics.) A
different value for $G_-$ and $G_+$ implies a different upstream half
width of the horseshoe region. Such a case is depicted in
Fig.~\ref{fig:scheme-asym}. The test particle $A$, which lies outside
of the separatrix, is circulating, and therefore mapped to $A'$ by the
flow. Similarly, the test particle $B$ lies inside of the separatrix,
and thus is librating. It is therefore mapped to $B'$ by the horseshoe
motion. There is therefore an over-density in the darker stripe in the
bottom right quadrant of Fig.~\ref{fig:scheme-asym}, as test particles
of different origins merge into this stripe. Reciprocally, a void
region appears at the downstream rear separatrix (the white stripe in
the top left quadrant of Fig.~\ref{fig:scheme-asym}).  This region can
neither be reached by the test particles executing horseshoe U-turns,
as these are too narrow, nor can it be reached by test particles of
the outer disk. The net effect of the asymmetry is that two stripes
of perturbed  surface density $\pm \Sigma_0$ appear at the 
downstream separatrices, with a width
\begin{equation}
  \label{eq:84}
  \delta x_s=\left|\frac{\Delta G}{\partial_rG}\right|=
\left|\frac{\Delta G}{4ABx_s}\right|
\end{equation}
The linear mass of these stripes is therefore, in absolute value:
\begin{equation}
  \label{eq:85}
  \lambda = \Sigma_0\left|\frac{\Delta G}{4ABx_s}\right|=\frac{\Sigma_0}
{2B}|\Delta v|=l_0|\Delta v|,
\end{equation}
where we have used Eq.~(\ref{eq:62}). This linear mass corresponds to
the factor of the $\delta$-function in the bracket of Eq.~(\ref{eq:81}).
The torque due to the stripes can be evaluated as follows. The
symmetric, barotropic case would be recovered if one sent the
excess of surface density of the overdense stripe to the empty
stripe, spending angular momentum for this purpose at the rate:
\begin{equation}
  \label{eq:86}
  |\Gamma_1|=4aBx_s\cdot |2Ax_s\Sigma_0\delta x_s|,
\end{equation}
where the factor in absolute value represents the mass flow rate in
the stripes. Using Eqs.~(\ref{eq:84}) and~(\ref{eq:86}), one recovers
the expression of the torque excess, given by the last term of the
right hand side of Eq.~(\ref{eq:37}).  In the case considered in
Fig.~\ref{fig:scheme-asym}, which has a wider horseshoe region at the
front of the planet, one would have to give angular momentum to the
material in excess in the front stripe, in order to recover a
symmetric situation. Differently said, this implies that in the
adiabatic situation the disk's material receives less angular momentum
than in the barotropic case, hence that there is a negative torque
excess on the disk, and a positive torque excess on the planet, by
virtue of the action and reaction law. This can also be deduced simply
from the distribution of perturbed density, since there is a positive
perturbed density in front of the planet and a negative one behind the
planet.

In a disk with pressure, narrow stripes of surface density are spread
radially by the excitation of evanescent waves,
which are not easily detectable (see
section~\ref{sec:press-dist}). They leave however an imprint on the
flow in the form of vorticity sheets. These are the most tangible
perturbations of the flow associated to the torque excess.

One can also understand that the torque excess due to these stripes
can in principle be extremely high. While the surface density
perturbation in the barotropic case typically amounts to
$O(\Sigma_0x_s/a)$ (prior to the convolution by the evanescent wave
kernel), the stripes contemplated here correspond to a perturbation of surface density that is
$\pm \Sigma_0$. If the asymmetry
becomes large enough to represent a sizable fraction of $x_s$, the
linear mass of the perturbation associated to the excess supersedes
the linear mass of the vortensity related perturbation by a factor
$O(a/x_s)$, that is to say typically by two orders of magnitude, for the
situations considered here.

\begin{figure}
  \centering
  \plotone{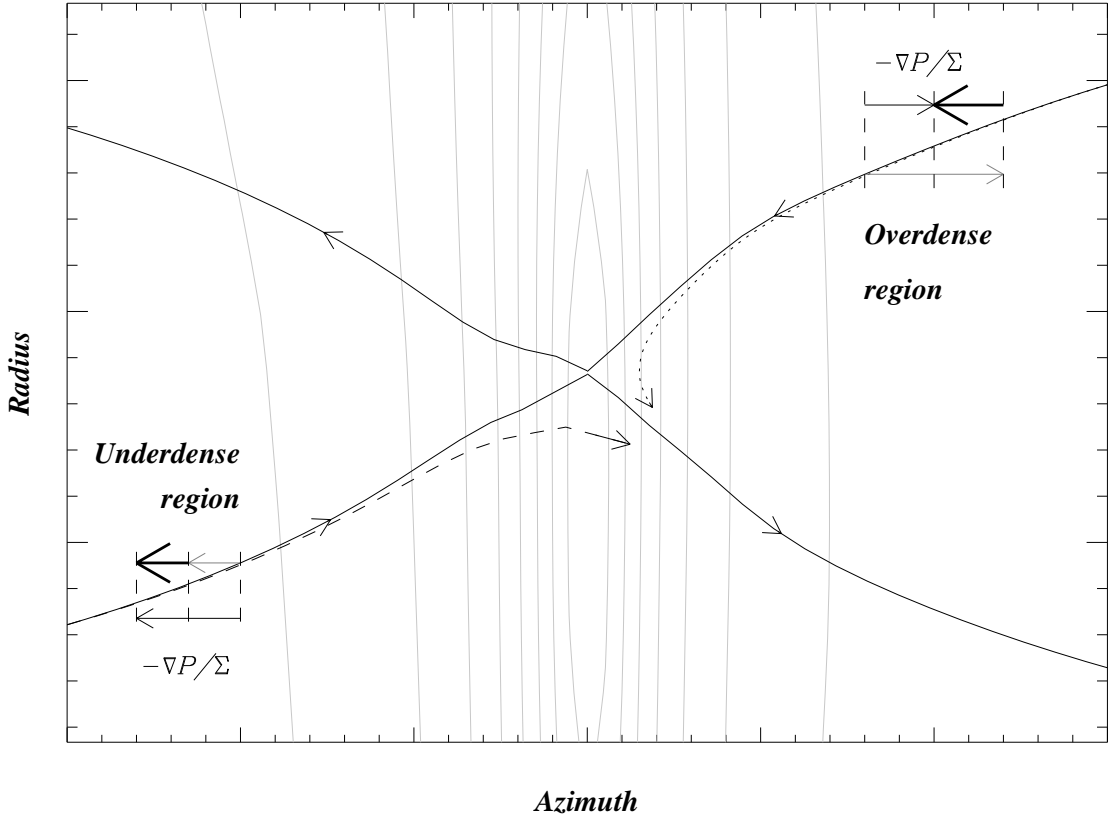}
  \caption{Schematic view of the action of the pressure gradient in an
    idealized case. The grey contours represent the perturbed
    pressure. The solid lines depict the separatrices of a barotropic
    case. The grey thin arrows represent $-\nabla P/\Sigma$ in the
    barotropic case, and the black thin arrows the same quantity in
    the adiabatic case, the difference arising essentially from the
    variations of $\Sigma$ between both cases. The thick arrows
    represent the excess of $-\nabla P/\Sigma$, which is negative on
    both sides of the planet. The fluid particle initially located on
    the rear upstream separatrix of the barotropic case therefore
    becomes circulating, as depicted by the dashed
    streamline. Similarly, the fluid parcel of the front separatrix
    loses more angular momentum than it would lose on the
    barotropic separatrix, hence the streamline associated to it is
    librating, as indicated by the dotted streamline.}
  \label{fig:scheme-ideal-presgrad}
\end{figure}

Further insight into the origin of the asymmetry can be gained
by considering, for the sake of simplicity, an idealized situation
in which the stagnation point lies at $\phi=0$, i.e. at the maximum
of the perturbed pressure. We can then evaluate, in order of
magnitude, how the torque acting on a fluid element moving
along an upstream separatrix differs from the barotropic case,
and infer from this how the separatrix should be shifted. This
situation is depicted in Fig.~\ref{fig:scheme-ideal-presgrad}.
For the sake of definiteness we assume that there is a negative
radial entropy gradient.

The perturbation of density associated to the advection of entropy
in the adiabatic case \citep[see][]{bm08} enhances or diminishes
the specific pressure torque $-\nabla P/\Sigma$. As can be seen
in Fig.~\ref{fig:scheme-ideal-presgrad}, a consequence of this
change is to render the horseshoe region asymmetric: the rear
part shrinks, whereas the front part expands. A coarse estimate
of the variation of the width of the horseshoe region can be
given as follows. A fluid element on a separatrix, when it reaches
the azimuth of the stagnation point, has received exactly the amount $J$
of angular momentum necessary to bring it to the stagnation point (by
definition of the separatrix):
\begin{equation}
  \label{eq:87}
  J=\int_{-\infty}^{t_s}\left(-\partial_\phi\Phi-\frac{\partial_\phi P}{\Sigma}
\right)dt=2Br_px_s,
\end{equation}
where $t_s$ is the date at which the fluid element reaches the
stagnation point. Assuming the fluid element to have at any instant
in time the azimuthal velocity of the unperturbed Keplerian shear,
we can write:
\begin{equation}
  \label{eq:88} J=\int_{-\infty}^{\phi_s}\frac{r_p}{2Ax}\left(\partial_\phi\Phi+\frac{\partial_\phi P}{\Sigma}
\right)dt=2Br_px_s.
\end{equation}
A rough estimate of how $x_s$ varies when the pressure term is
modified can be obtained by taking $x$ out the integral in
Eq.~(\ref{eq:88}), assuming that it is everywhere equal to $x_s$. At
this level of approximation, the perturbation of the pressure term can
be written as $\sim (\partial_\phi P/\Sigma_0)(2{\cal S}x_s/a)$. We are
then left with:
\begin{equation}
  \label{eq:89}
  \left(\int_{-\infty}^{\phi_s}\frac{\partial_\phi P}{\Sigma_0}d\phi\right)
.(2{\cal S}x_s/a) \sim 2ABx_s\delta x_s,
\end{equation}
which can be recast as:
\begin{equation}
  \label{eq:90}
  \delta x_s\sim \frac{p_s{\cal S}}{2ABa\Sigma_0}.
\end{equation}
We note that we obtain the same expression for the variation of the
horseshoe width using Eqs.~(\ref{eq:41}), (\ref{eq:50}), (\ref{eq:54})
and (\ref{eq:84}), within a factor of order unity (the value $\delta
x_s$ obtained above just refers to the variation of the width of one
side of the horseshoe region, whereas the expression worked out
earlier in this section and depicted by the thin double arrows in
Fig.~\ref{fig:scheme-asym} corresponds to adding the width variation
of the front and rear sides.)

This order of magnitude estimate allows to identify the origin of the
horseshoe asymmetry, and allows to understand why the adiabatic
torque excess scales with the perturbation of pressure at the
stagnation point ($p_s$), at least in the regime of small planetary masses
(for which $p_s\ll P_0$.) We finally note that, while this order
of magnitude was obtained in the simplified case of a stagnation point
located at the maximum of perturbed pressure, it still holds when
the stagnation point is azimuthally shifted, as in real situations.
A more refined treatment of the dynamics of a fluid element moving
along the separatrices would have exhibited its Bernoulli invariant,
and it would have led to an exact relationship such as Eq.~(\ref{eq:44}), which is equivalent to Eq.~(\ref{eq:88}). This 
relationship shows that the width of the horseshoe region depends
exclusively of the flow properties of the stagnation point, independently
of the location of the latter and of the path followed by the fluid
elements.

\section{A suitable torque expression}
\label{sec:suit-torq-expr}
The torque expression of Eq.~(\ref{eq:55}) involves the planetary
potential at the stagnation point, as well as the distance of the
upstream separatrices to corotation.  As such, it is not well suited
for torque estimates, and needs further transformation.

We have seen in section~\ref{sec:note-half-width} that the half width
of the horseshoe region has a complex dependence on the disk's and
planet's parameters in the adiabatic case.  In particular, at large
values of $|{\cal S}|$, the horseshoe region is wider than in the
barotropic case, which boosts the vortensity related corotation
torque. The global torque excess is therefore the sum of two
non-trivial individual excesses: the one related to the gradient of
entropy, which has been studied in depth in the preceding sections,
and the one related to the boost of the vortensity related corotation
torque. We note however in Fig.~\ref{fig:excess-entr} that the global
excess, which corresponds to the dotted curve, has an almost linear
dependence on the entropy gradient. Our aim is therefore to exhibit a
separable form of the torque excess that involves the product of
${\cal S}$ by a function of $\hat\epsilon$. This expression aims
at describing the global torque excess, therefore accounting for the
whole difference between the adiabatic and barotropic case. For this
purpose we can simply use the data of the series SA$_i$, which has a
null vortensity gradient and for which the adiabatic torque excess
accounts for the whole excess. We then rescale the function obtained
to get the excess at any entropy gradient.

\subsection{Scaling of the torque excess}
We first establish the scaling of the torque excess as a function of the
disk's and planet's parameters. Eq.~(\ref{eq:55}) shows that the
torque excess involves the half width of the horseshoe region, and
the potential at the stagnation point. In order to remain in the framework
of a separable expression, we neglect the dependence of $x_s$ on the
entropy gradient, mentioned in section~\ref{sec:note-half-width}, and
we use the barotropic approximation (i.e. for ${\cal S}=0$) to infer
the half width. Using Eq.~(\ref{eq:57}), we obtain:
\begin{equation}
  \label{eq:91}
  x_s^{\rm adi}=\frac{C(\hat\epsilon)}{\gamma^{1/4}}a\sqrt\frac qh,
\end{equation}
where $C(\hat\epsilon)$ is a dimensionless constant that depends on
the softening length \citep{2009arXiv0901.2263P}.
Examination of Fig.~\ref{fig:hwhs} shows that this approximation leads
to errors of at most $\sim 30$~\%. We check in what follows
that the potential at the stagnation point can also be considered as
essentially independent of the entropy gradient.

\subsubsection{Potential at the stagnation point}
\label{sec:potent-at-stagn}
Evaluating the planetary potential at the stagnation point requires to
know the location of this point.  Writing the Euler equations in
steady state, in a system of units in which the unit of length is
$H=c_s/\Omega$, one can realize that, as long as the planet is deeply
embedded (i.e. $aq^{1/3}\ll H$), the azimuth of the stagnation point
(which lies almost on corotation) has to scale as $hf(\hat\epsilon)$,
where $f$ is a function that we shall determine. We have therefore:
\begin{equation}
  \label{eq:92}
  \Phi_p ^s=-\frac{GM_p}{H[f(\hat\epsilon)^2+\hat\epsilon^2]^{1/2}}.
\end{equation}
\begin{figure}
  \centering
  \plotone{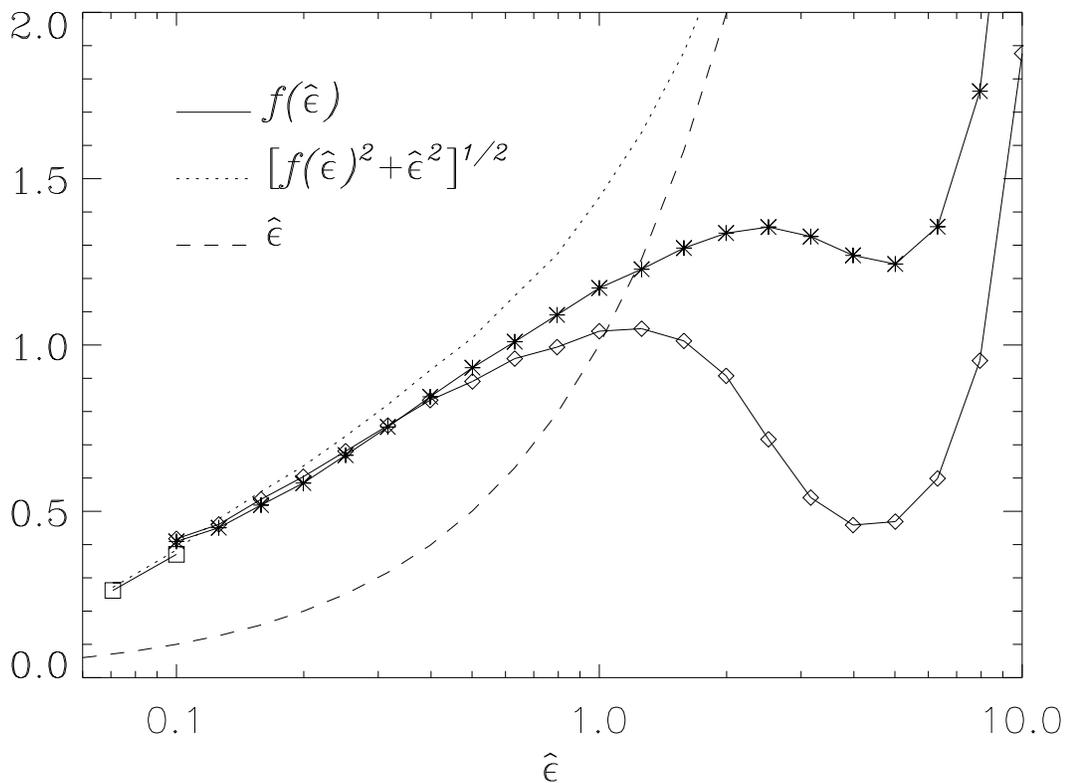}
  \caption{Location of the stagnation point as a function of the
    softening length, measured at $t=40$~orbits. Diamonds show
values obtained in the EA$_{0\le i\le 20}$ series. The two squares show
values obtained for the same parameters as in the EA series, except that
the resolution has been increased to $2000\times 2000$. The stars show
values obtained from a series similar to EA, except that it has a different
entropy gradient: ${\cal S}=+12/7$. The dotted curve, which shows
$[f(\hat\epsilon)^2+\hat\epsilon^2]^{1/2}$, has been obtained using
the data provided by the EA series.}
  \label{fig:stag}
\end{figure}
Fig.~\ref{fig:stag} shows the location of the stagnation point as a
function of $\hat\epsilon$ in an adiabatic flow with $\gamma=1.4$,
obtained from the series of numerical simulations EA$_{0\le i\le 20}$
(see section~\ref{sec:numerical-set-up}), and a similar series with a
different entropy gradient. Under $\hat\epsilon\simeq 1.0$, the
distance of the stagnation point to the planet is larger than the
softening length, and the denominator of Eq.~(\ref{eq:92}) is
dominated by the function $f(\hat\epsilon)$, which is found to
hardly depend on ${\cal S}$, as both series essentially coincide
for $\hat\epsilon < 1.0$. Reciprocally, for
$\hat\epsilon>1.0$, the stagnation point lies within a softening
length from the planet, and the potential at the stagnation point
approximately scales as $\hat\epsilon^{-1}$, and is therefore also
independent of the entropy gradient.

\subsubsection{General expression of the torque excess}
Using Eqs.~(\ref{eq:55}), (\ref{eq:91}) and (\ref{eq:92}), we obtain:
\begin{equation}
  \label{eq:93}
  \Gamma_1=-A(\hat\epsilon){\cal S}\Gamma_0,
\end{equation}
where $\Gamma_0=\Sigma_0\Omega_p^2q^2a^4/h^2$, and where:
\begin{equation}
  \label{eq:94}
  A(\hat\epsilon) = \frac{4C^2(\hat\epsilon)}{\sqrt {1.4}}\left[\frac{1}{\sqrt{f(\hat\epsilon)^2+\hat\epsilon^2}}-\frac 38\frac{C^2(\hat\epsilon)}{\sqrt{1.4}}\right],
\end{equation}
where we have specialized to the Keplerian case ($A=-3\Omega/4$,
$B=\Omega/4$). Note that we explicitly restrict ourselves to the case
$\gamma=1.4$. Although Eq.~(\ref{eq:91}) applies for arbitrary values
of $\gamma$, the azimuth of the stagnation
point, considered in section~\ref{sec:potent-at-stagn}, may depend 
on the adiabatic index. Keeping an explicit dependence
in $\gamma$ in Eq.~(\ref{eq:94}) could therefore be misleading.  The
adiabatic torque excess, given by Eq.~(\ref{eq:93}), has a scaling
similar to the barotropic corotation torque or the differential
Lindblad torque \citep{tanaka2002}. The only differences with the barotropic
corotation torque is that the excess scales with ${\cal S}$ rather
than ${\cal V}$, and that it involves a different dimensionless
function of $\hat\epsilon$.

In order to determine this function, we use the series SA$_i$, as
explained above. We restrict ourselves to the range
$0.1\le\hat\epsilon\le1$.  Values of $\hat\epsilon$ larger than one
are not relevant, whereas values smaller than $0.1$ are difficult to
study owing to the very high resolution required. As mentioned in section~\ref{sec:depend-soft-length}, the torque excess exhibits a
dependence in $\hat\epsilon^{-1}$ with a good approximation over
the range of $\hat\epsilon$ considered (a dependence partly due to
the increase of $x_s$ at small softening length, and partly due to the
increase of the planetary potential at the stagnation point at small
softening length). Fitting the torque excess over the range of $\hat\epsilon$
considered, we obtain:
\begin{equation}
  \label{eq:95}
  A(\hat\epsilon) \approx \frac{1.3}{\hat\epsilon}.
\end{equation}
The expression of the torque excess is therefore:
\begin{equation}
  \label{eq:96}
  \Gamma_1=-\frac{1.3{\cal S}}{\hat\epsilon}\Sigma_0\Omega_p^2q^2a^4h^{-2}.
\end{equation}
One can see in Figs.~\ref{fig:excess-smoo} and~\ref{fig:excess-entr}
that this expression (which is represented by the grey dashed line) is
in good agreement with the global adiabatic excess. It is also in good
agreement with the results of \citet{bm08}.

\subsection{Considerations on the three-dimensional case}

If one considers that a two-dimensional case with softening length
$\epsilon$ represents the layer at altitude $z=\pm\epsilon$ of a three
dimensional case \citep[see e.g.][]{masset02}, one can perform an
integration of the torque over $\epsilon$ in order to get an
approximate value of the three dimensional torque. A layer comprised
between altitude $z$ and $z+dz$ has a surface density
$(\Sigma_0/\sqrt{2\pi}H)\exp[-z^2/(2H^2)]dz$, hence the three
dimensional adiabatic torque excess should be approximately given by:
\begin{eqnarray}
  \label{eq:97}
  \Gamma_{3D}&\approx& -2\int_0^{+\infty}\frac{\Sigma_0}{\sqrt{2\pi}H}\exp[-z^2/(2H^2)]
A\left(\frac zH\right){\cal S}\Omega_p^2\left(\frac qh\right)^2r_p^4dz\nonumber\\
&\approx&-\sqrt\frac 2\pi{\cal S}\Gamma_0\int_0^{+\infty}e^{-\hat\epsilon^2/2}
A(\hat\epsilon)d\hat\epsilon
\end{eqnarray}

While this kind of calculation can be performed relatively easily for
the vortensity related torque (for which the dimensionless function of
$\hat\epsilon$ converges when $\hat\epsilon\rightarrow 0$) or for the
differential Lindblad torque, it is a risky exercise in the case of
the adiabatic torque excess, since the approximate expression of
Eq.~(\ref{eq:95}) diverges for $\hat\epsilon\rightarrow 0$.  For the
setup that we consider, we have tried to perform calculations at
higher resolution and smaller smoothing lengths. The smallest
smoothing length for which we have a steady situation is $\epsilon =
0.07H$ (corresponding to the left square in
Fig.~\ref{fig:stag}). Under this value, we end up with very messy,
unsteady situations, in which the flow is strongly affected by the
presence of numerous vortices which drift along the separatrices.
We can therefore only provide a conservative estimate of Eq.~(\ref{eq:97}),
by truncating the integral at $\hat\epsilon=0.07$. We obtain:
\begin{equation}
  \label{eq:98}
  \Gamma_1^{3D}\approx -2.8{\cal S}\Gamma_0.
\end{equation}
In order to estimate the total torque in the three-dimensional case, we
need to know the differential Lindblad torque, and the vortensity related
part of the horseshoe drag. The first of those is given by \citet{tanaka2002}, and
is, with our notation:
\begin{equation}
  \label{eq:99}
  \Gamma_{LR}=-(2.34-0.1\alpha)\Gamma_0,
\end{equation}
for a disk without a temperature gradient.
We note that we do not use the total torque value given by \citet{tanaka2002},
but only the differential Lindblad torque. The total torque value contains indeed
the linear corotation torque, which is not the corotation torque exerted 
in steady state \citep{2009arXiv0901.2265P}. Instead, we consider separately
the vortensity related horseshoe drag, and apply to it a treatment similar to
that of Eq.~(\ref{eq:97}), which is straightforward in this case, because of the
absence of divergence when $\hat\epsilon\rightarrow 0$. The two-dimensional
component of the horseshoe drag that scales with the vortensity gradient reads
\citep{wlpi91,masset01}:
\begin{equation}
  \label{eq:100}
  \Gamma_V=\frac34{\cal V}C^4(\hat\epsilon)\Gamma_0,
\end{equation}
hence its three-dimensional estimate is:
\begin{equation}
  \label{eq:101}
  \Gamma_V^{3D}\approx\frac{3}{\sqrt{8\pi}}{\cal V}\Gamma_0\int_0^{+\infty}e^{-\hat\epsilon^2/2}
  C^4(\hat\epsilon)d\hat\epsilon\approx 0.93{\cal V}\Gamma_0,
\end{equation}
where we have used the series SI$_i$ to tabulate $C(\hat\epsilon)$.
The total torque acting on a low-mass planet in a three-dimensional disk
reads therefore:
\begin{equation}
  \label{eq:102}
  \Gamma=[-(2.34-0.1\alpha)+0.93{\cal V}-2.8{\cal S}]\Gamma_0,
\end{equation}
where we recall that the last term is a conservative estimate, and is likely
to actually have a numerical coefficient larger than $2.8$. We see that
for $|{\cal S}|\sim|{\cal V}|\sim 1$, the term associated to the entropy
gradient dominates the horseshoe drag, hence the adiabatic torque excess
should efficiently halt or reverse migration in disks with moderate, negative
entropy gradients, in agreement with the result of \citet{pm06}, who
found that planetary migration can indeed be reversed in a three-dimensional
disk.

\section{Discussion}
\label{sec:discussion}
In this section we describe some side results, and we draw a list
of points that deserve further investigation.

\subsection{Topology of the flow and location of the stagnation point}
\label{sec:topol-flow-locat}
As stated in section~\ref{sec:assumptions}, we have assumed in this
work that there is only one X-point in the vicinity of the
planet. We find that this is indeed
the case in most situations. We can nevertheless get two X-points
in the vicinity of the planet, if one of the following conditions is
fulfilled:
\begin{itemize}
\item the entropy gradient vanishes, or is very small,
\item the softening length of the potential is very small,
\item the flow has not reached a steady state and is observed
  before a U-turn time scale.
\end{itemize}
While the first and last conditions are not relevant to the present
work, which assumes a steady state and a sizable entropy gradient, 
the second case has some importance, as it gives indications of what
can happen near the equatorial plane in a three-dimensional case.
In the cases in which we have two X-points that subsist during the
whole simulation, we get very time-dependent torque values. This
can be interpreted as entropy trapped in the small libration island
defined by the two X-points. The perturbation of density associated
to the entropy distribution, that librates in this island, gives rise
to large torque variation, as it corresponds to material very close
to the planet. We note that this is quite in contrast with what we have claimed
in section~\ref{sec:press-dist}, where we have argued that it is the
vortensity perturbation that  yields the bulk torque. Whereas this is
true for the downstream horseshoe stripes which lie at some distance
of the planet, here the planet lies within the libration island, and the
fact that the density response is spread radially or not definitely matters.
We also note that the present discussion offers a number of similarities
with the discussion in paper~I about the flow topology as a function
of the temperature gradient and  of the softening length, except that
here it is the entropy gradient that plays a role, rather than the temperature gradient.

When the flow is observed at an early stage, it may display two
X-stagnation points, which are generally not on the same
separatrix. We observe that once the flow has settled in a steady
state, it is the outermost stagnation point (the one that is on the
separatrix that lies further from corotation) that subsists. It turns
out that this point lies behind the planet for ${\cal S}<0$, and is in
front of the planet for ${\cal S}>0$. While this fact is unimportant
for a planet in fixed circular orbit, where only the distance of
the stagnation point to the planet matters to evaluate the torque
excess, it is crucial when the planet migrates, as the stagnation point
is shifted from the position that it has on a fixed circular orbit.
The consequences of this are discussed in the next section.

\subsection{Feedback on migration}
\label{sec:feedback-migration}
\begin{figure}
  \centering
  \plotone{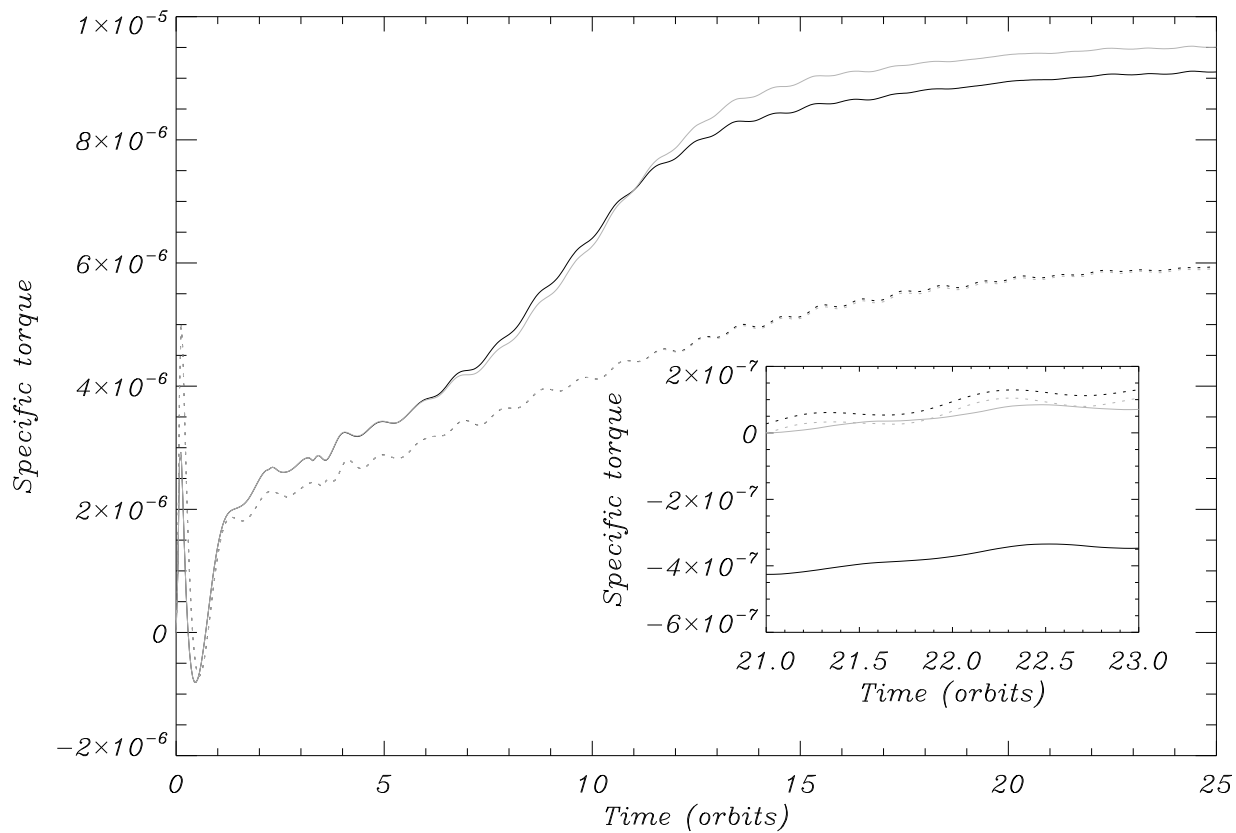}
  \caption{Specific torque on planet in the runs with an inwards disk drift. The solid
line shows the adiabatic case and the dotted line shows the isothermal case. Grey curves
show the initial runs (without disk drift). The inset plot shows a close up on the
curves (the value of the torque at $t=21$~orbits in the case without disk drift has been
subtracted to improve legibility).}
  \label{fig:drift}
\end{figure}

\begin{figure}
  \centering
  \plotone{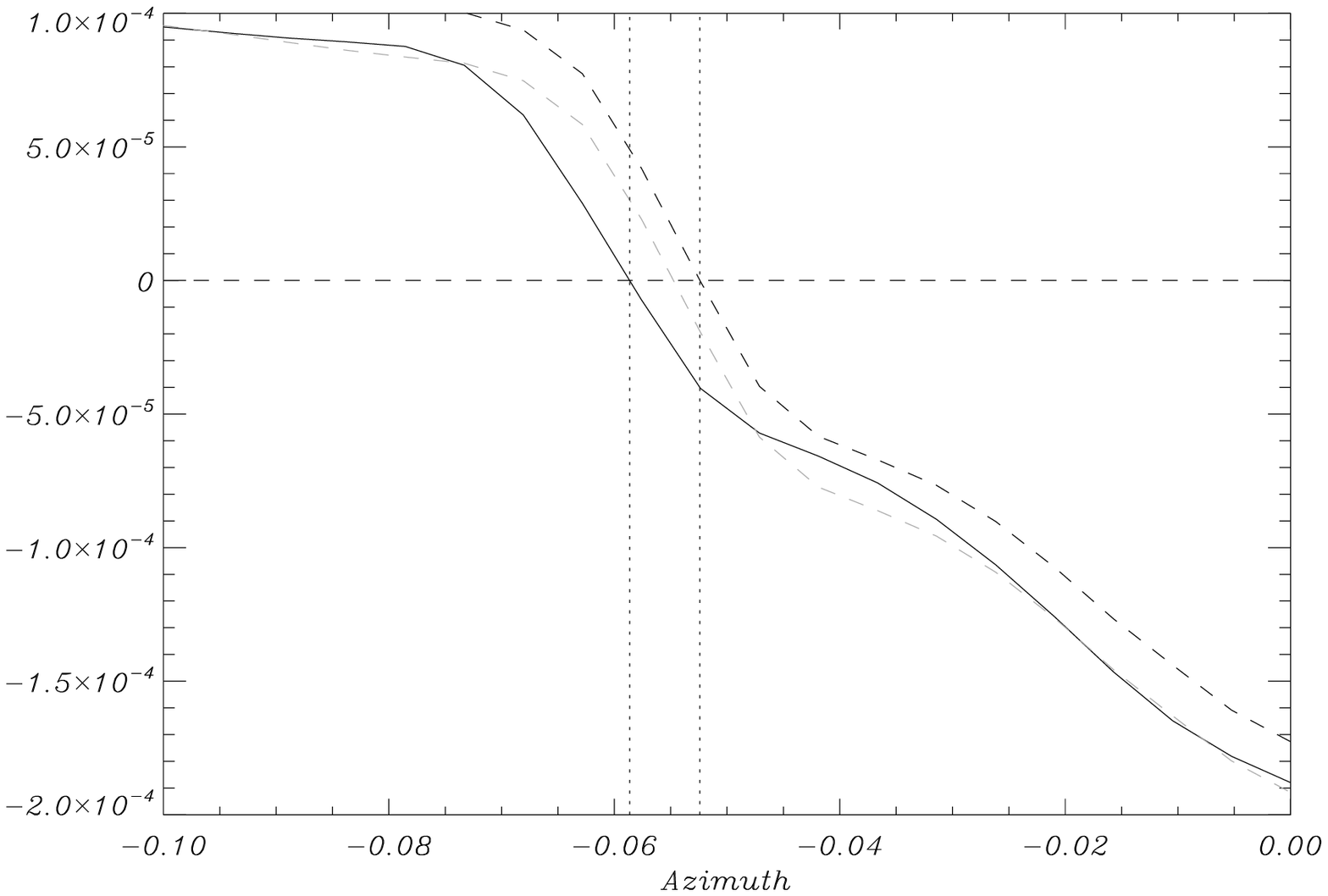}
  \caption{Radial velocity as a function of the azimuth, at $r=r_c$,
    in the static case (dashed line), and in the case with disk drift
    (solid line). The grey dashed curve shows the curve of the static
    case offset of $\dot r_d$. The rightmost vertical dotted line
    shows the initial location of the stagnation point, while the
    leftmost one shows the location of the stagnation point in the
    case with a disk drift, which is also the location expected for a
    planet moving outwards in the non drifting disk. }
  \label{fig:vrad-drift}
\end{figure}

Since the adiabatic torque excess depends on the flow properties at
the stagnation point of the horseshoe region, and since, as we shall
see below, the location of the latter depends on the drift rate, we
expect that a feed back loop can be established between the total
torque and the migration rate, much as in type~III migration
\citep{mp03}.  Contrary to type~III migration, however, the feed back
should generally be negative, therefore acting to decrease the
absolute value of the drift rate. It should also be of relatively
minor importance, except in very massive disks, with Toomre's $Q$
parameters close to unity. Consider the following numerical
experiment: we repeat the runs SA$_{10}$ and SI$_{10}$, except that we
impose an inwards disk drift $\dot r_d=-1.9\cdot 10^{-5}$. We note
that, if the planet was allowed to freely migrate in run SA$_{10}$, it
would do so at the rate $\dot a=1.9\cdot 10^{-5}$. Rather than
releasing the planet, we impose an inwards drift of the disk, of same
magnitude. This technique has already been used for type~III migration
\citep[see][section~5.6]{mp03}. We see in Fig.~\ref{fig:drift} that
the torque in the isothermal case does not depend on the drift rate,
while there is a noticeable difference in the adiabatic case, and that
the torque tends towards a somehow smaller value. The reason for this
can be understood from the profile of radial velocity at corotation,
depicted in Fig.~\ref{fig:vrad-drift}. If the planet migrates
outwards, the stagnation point (in a frame that moves radially with
the planet) is located where the horseshoe dynamics endows fluid
elements with a radial velocity equal to that of the planet
(similarly, if the planet is kept fixed, the new stagnation point is
expected to lie where the radial velocity in the initial case is
$-\dot r_d$.) Since fluid elements move outwards behind the stagnation
point, the azimuth of the stagnation point decreases, hence the latter
recedes from the planet, since it is located behind the planet in the
case considered. A consequence of this recession is a lower perturbed
pressure at the new position of the stagnation point, hence a lower
value of the adiabatic torque excess. A similar decrease (in absolute
value) is also expected in the case ${\cal S}>0$. The stagnation point
is then in the front of the planet (see
section~\ref{sec:topol-flow-locat}), and the total torque acting on
the latter is negative. The stagnation point in the migrating case is
therefore expected where the horseshoe U-turns are performed inwards,
that is to say at the front of the stagnation point of the fixed case,
and therefore further again from the planet. We have checked this with
additional simulations, not reproduced here.

A number of comments are in order:
\begin{itemize}
\item Fig.~\ref{fig:vrad-drift} shows a poor agreement between the radial
velocity for the disk's drift case and the original radial velocity offset by
$-\dot r_d$, in the region of the stagnation point ($-0.07\le\phi\le-0.03$).
This is presumably due, at least in part, to the resolution of the grid. The
direction of the shift of the stagnation point, and its order of magnitude, are
nevertheless compatible with the new stagnation point being determined by
the location where, in the initial flow, $v_r=\dot a$.
\item Since both $v_r$ and $\dot a$ scale with the planet mass (for the range
of planetary masses considered here), the shift of the stagnation point should not
depend on the planetary mass.
\item Since $\dot a$ scales with the disk mass, the shift of the
  stagnation point of a freely migrating planet should increase with
  the disk mass. The magnitude of this effect, for the disk considered in the
numerical experiment described above, is at about $1/20^{th}$ of  the total torque.
The disk has a Toomre parameter $Q \approx 8$. This shows that the effect we
describe is of marginal importance, except in the most massive disks, for which
$Q\gtrsim 1$.
\item The drift rate in steady state is given by:
  \begin{equation}
    \label{eq:103}
    2Ba\dot a=\gamma+\frac{\partial\gamma}{\partial\dot a}\dot a,
  \end{equation}
  where we assume that the torque excess, to lowest order, has an
  affine dependence on $\dot a$, and where $\gamma$ is the specific
  torque acting on the planet.  We therefore obtain:
\begin{equation}
  \label{eq:104}
  \dot a=\frac{\gamma}{2Ba-\partial\gamma/\partial\dot a}.
\end{equation}
In the particular case described above, we have
$\partial\gamma/\partial\dot a \approx -0.05 \cdot 2Ba$, hence the
steady state drift is almost equal to that dictated by the torque
value measured in the disk drift case (i.e. further iterations with a
drift rate of the disk given by the new torque value would hardly change the result).
\end{itemize}
 
\subsection{Extension to arbitrary temperature profiles}
\label{sec:extens-arbitr-temp}
The analysis presented in this work suffers from a restriction very
similar to the restriction of the isothermal case, which is that the
temperature profile is assumed to be flat. We have nevertheless
performed many additional calculations in which we relaxed this
constraint, and found that our main results essentially hold in disks
with arbitrary temperature profiles:
\begin{itemize}
\item The global torque excess, defined by Eq.~(\ref{eq:57}), scales
  with the entropy gradient.  The isothermal runs which we use to
  apply Eq.~(\ref{eq:57}) in the general case have same profiles as
  their adiabatic counterparts, hence they are locally isothermal
  runs, which have singular evanescent waves excited at the
  separatrices, contrary to the globally isothermal runs (see
  paper~I). These components, which scale with the temperature
  gradient, should therefore be also present in the general, adiabatic
  case.
\item The half width of the horseshoe region exhibits the same characteristic V-shaped
dependence on ${\cal S}$, as shown in Fig.~\ref{fig:hwhs}.
\end{itemize}
The adiabatic torque excess, as a function of the entropy gradient,
displays little scatter, and is therefore essentially a one-to-one
correspondence. The results presented here can therefore be used to
predict the torque in a general situation. They
indicate that, in the general case, the horseshoe drag consists
of three terms: the classical vortensity related horseshoe drag, 
the adiabatic torque excess presented in this work, and an additional
term which scales with the temperature gradient (see paper~I). The
latter corresponds to the creation of vortensity in the vicinity of the
planet, near the stagnation point, where the flow is slow and the driving
of vortensity is efficient. We comment that the adiabatic torque excess can also be
seen as a singular creation of vortensity at the downstream separatrices,
with the word of caution that vortensity is mathematically ill-defined at these separatrices,
because of the presence of a contact discontinuity (so that one has to consider
the low order estimate $\omega/\Sigma_0$, rather than $\omega/\Sigma$).

\subsection{Adiabatic excess in turbulent disks}
The analysis presented above relies upon the existence of a steady
state and on that of a unique stagnation point. It therefore does not
apply to the case of a turbulent disk. The adiabatic excess has been
interpreted, however, as due to a constitutive asymmetry of the
horsehoe region, acquired by the upstream flow under the action of the
pressure gradient on overdense or underdense regions. To some extent,
this effect should therefore persist in a turbulent disk, since we
expect to find, in a turbulent disk, regions that are, in average,
under- or over-dense with respect to the barotropic case, as a
consequence of the advection of entropy. This issue deserves
significant further work, since most of the regions of the disks where
planetary migration is supposed to take place should be turbulent. In
particular, it would be of interest to know whether the time-averaged
torque excess is equal to the steady state estimate.

\subsection{Saturation issues}
In the absence of any process that can transfer angular momentum
between the horseshoe region and the rest of the disk, the horseshoe
drag is bound to saturate. A non-vanishing time averaged value of this
torque would indeed imply a sustained transfer of angular momentum
from the planet to the horseshoe region, where it would accumulate.
Viscous torques, exerted at the separatrices between the horseshoe
region and the outer or inner disk, can ensure this transfer of
angular momentum \citep{masset01}. In the case of an isothermal disk,
viscous diffusion tends to restore the large scale gradient of
vortensity across the horseshoe region. If it can do so in less than a
libration time, the vortensity upstream of the horseshoe U-turn is
that of the unperturbed disk, and the horseshoe drag is permanently
equal to its initial value.  In the case of an adiabatic flow, the
vortensity related part of the horseshoe drag follows the same
picture, and a sufficiently large viscosity is required to avoid the
saturation of this component of the drag. The adiabatic torque excess,
however, results from the advection of entropy, and to the existence,
at the stagnation point, of a discontinuity of the entropy brought
from the inside and from the outside. Sustaining the torque excess
therefore implies that the discontinuity of entropy at the stagnation
point is maintained.  \citet{pp08} consider thermal diffusion as the
dissipative process that fulfills this function, while
\citet{2008A&A...487L...9K} consider realistic radiative effects. In
both cases there can be a sustained, positive torque acting on the
planet.  We note however:
\begin{itemize}
\item that in any case a finite amount of viscosity is
  required. Thermal diffusion or radiative effects, which do not
  feature in Euler's equation, cannot ensure the transfer of angular
  momentum out of the horsehoe region.
\item That there is no theoretical expression of the steady state
  horseshoe drag that takes into account the balance between
  saturation and the dissipative processes that prevent it. Such an
  expression is definitely required for planet population synthesis
  based on one-dimensional models of migration.
\end{itemize}

We also comment that, while saturation can be interpreted as the
phase mixing of the contribution of different horseshoe streamlines in
the barotropic case \citep{bk2001}, the case of the adiabatic torque
excess is slightly different because there is essentially one
streamline involved in this excess: the separatrix. It is therefore
possible that the torque excess oscillates over a very long time, or
indefinitely.  Numerical simulations show that the oscillations of the
torque damp after a few libration times \citep{phdbaruteau}. It is not clear, however,
whether this is a physical effect or the consequence of numerical
diffusivity. This again appeals for the need of simulations at very
high resolution.

We finally mention the systematic appearance of a vortex at one of the
downstream separatrices (the one that has a negative singularity of
vorticity). It would be of interest to study the impact of this vortex
on the saturation of the adiabatic excess. On the one hand, it
provides a mixing of the entropy in the vicinity of the separatrix,
and on the other hand, it may assist the horseshoe region in
exchanging angular momentum with the rest of the disk.

\section{Conclusions}
\label{sec:conclusions}
We have shown that the horseshoe drag exerted on a low-mass planet,
embedded in a disk that behaves adiabatically on the time scale of
a horseshoe U-turn, can be considered as the sum of two terms:
\begin{itemize}
\item a bulk term, involving all the horseshoe streamlines, which
  necessarily scales with the vortensity gradient, exactly as in a
  barotropic disk. This is in agreement with the forecast made in the
  introduction: there is no room, for a generic horseshoe streamline,
  for an exchange of angular momentum with the planet different from
  that of a barotropic situation.
\item An edge term, which manifests itself as a torque excess with
  respect to the barotropic case. This term scales with the radial
  gradient of entropy, and it corresponds to a constitutive asymmetry
  of the horseshoe region. In a pressureless disk, this asymmetry would trigger the appearance of underdense
  or overdense thin stripes of gas at the downstream
  separatrices. In a gaseous disk, pressure forces spread radially these stripes by the
  excitation of evanescent waves, leaving imprints as vorticity sheets
  at the separatrices as the most tangible manifestation of this
  effect.
\end{itemize}
We provide an expression for the torque excess, given by Eq.~(\ref{eq:96}),
which should be added to the differential Lindblad torque and to the
vortensity related horseshoe drag whenever the flow behaves adiabatically on
the timescale of the horseshoe U-turns. We also provide a tentative expression
of the total torque in the three-dimensional case, given by Eq.~(\ref{eq:102}).

We find that the origin of the adiabatic torque excess is not the
conspicuous under- or over-dense regions that appear within the
horseshoe region and which are bound by a contact discontinuity at the
downstream separatrices, as originally thought. Instead it is due to
rather discreet single evanescent waves launched at the downstream
separatrices. This answers some issues arising from earlier works. In
particular, no adiabatic torque excess is expected in the limit of a
cold disk \citep{bm08}, whereas the contact discontinuities still
exist in this limit and delineate perturbations of finite mass. The
torque excess that we find scales with the perturbed pressure at the
horseshoe's stagnation point, hence it vanishes in the limit of a cold
disk.

We found a number of side results:
\begin{itemize}
\item The horseshoe region in the adiabatic case has a more complex
  dependence on the disk's parameters than in the barotropic case. In
  particular, the horseshoe region is wider than expected from
  barotropic estimates, when there is a non-vanishing entropy
  gradient.  A consequence of this effect is that the bulk horseshoe
  term, that scales with the vortensity gradient, is boosted with
  respect to its value in a barotropic disk. This effect is relatively
  weak, however, and most of the difference between the adiabatic and
  barotropic case comes from the edge term, in disks with realistic
  profiles.
\item The adiabatic torque excess slightly depends on the migration
  rate.  There is therefore a weak feed back on migration. This feed
  back is found to be generally negative (i.e. it tends to lower the
  drift rate, either inward or outward), and is virtually negligible
  except in very massive disks.
\end{itemize}
We find that even at relatively high resolution, numerical simulations may
mistakingly locate the stagnation point, which has some impact on the 
torque excess value. We therefore stress the need for very high resolution
calculations, for which nested mesh codes would be a valuable tool. In a similar
vein, the steep dependence of the excess on the potential's softening length
suggests that the effect can be very strong in the three dimensional case. This
issue requires significant further work.

\acknowledgments

The numerical simulations performed in this work have been run on the
92 core cluster funded by the program {\em Origine des Plan\`etes et
  de la Vie} of the French {\em Institut National des Sciences de
  l'Univers}. Partial support from the COAST project ({\em COmputational
  ASTrophysics}) of the CEA is also acknowledged. The authors also wish
to thank G. Koenigsberger for hospitality at the Instituto de Ciencias Fisicas
of UNAM, Mexico, and acknowledge partial support from CONACYT project
number 24936. The authors are grateful to S.~Fromang and C.~Baruteau for
a thorough reading of a first draft of this manuscript.

\appendix

\section{Expression of the pressure perturbation}
\label{sec:expr-press-pert}
The convolution product of the rectangular function 
\begin{eqnarray}
  \label{eq:105}
  R:x\mapsto& 1\mbox{~if $0<x<x_0$}\\
  &0 \mbox{~otherwise,}\nonumber
\end{eqnarray}
by the kernel $K(x)=\exp(-|x|/\lambda)/(2\lambda)$ is:
\begin{eqnarray}
  \label{eq:106}
  \tilde R(x,x_0,\lambda) &=&1-\exp\left(-\frac{x_0}{2\lambda}\right)\cosh
\left(\frac{x-x_0/2}{\lambda}\right)\mbox{~if $x \in [0,x_0]$,}\\
&&\exp\left(-\frac{x-x_0/2}{\lambda}\right)\sinh\left(\frac{x_0}{2\lambda}\right)
\mbox{~otherwise.}\nonumber
\end{eqnarray}
Similarly, the convolution product of the triangular function 
\begin{eqnarray}
  \label{eq:107}
  T:x\mapsto& x\mbox{~if $0<x<x_0$}\\
  &0 \mbox{~otherwise,}\nonumber
\end{eqnarray}
by the kernel $K(x)=\exp(-|x|/\lambda)/(2\lambda)$ is:
\begin{eqnarray}
  \label{eq:108}
  \tilde T(x,x_0,\lambda) &=&\frac{\exp(x/\lambda )}{2}\left[
-(x_0+\lambda)\exp\left(-\frac{x_0}{\lambda}\right)+\lambda\right]\mbox{~if $x<0$,}\\
&&\frac{\exp(-x/\lambda )}{2}\left[
(x_0-\lambda)\exp\left(\frac{x_0}{\lambda}\right)+\lambda\right]\mbox{~if $x>x_0$,}\nonumber\\
&&x+\frac{\lambda}{2}\exp\left(-\frac x\lambda\right)-\frac{x_0+\lambda}{2}
\exp\left(\frac{x-x_0}{\lambda}\right)\mbox{~if $x\in [0,x_0]$.}\nonumber
\end{eqnarray}
Using Eq.~(\ref{eq:81}), we obtain:
\begin{eqnarray}
  \label{eq:109}
  \delta P&=&-\frac{\Gamma_1c_s^2}{(8a|A|Bx_s^2).(2\gamma^{1/2}H)} \exp\left(-\frac{|x|}{\sqrt\gamma H}\right)
-\frac{2\Sigma_0c_s^2}{r_p}({\cal S}+{\cal V})\tilde T(x,x_s,\sqrt{\gamma }H)\\
&&+\Sigma_0c_s^2\frac{\delta l_{\rm prod}}{l_0}\tilde R(x,x_s,\sqrt{\gamma} H),\nonumber
\end{eqnarray}
where we have also made use of Eq.~(\ref{eq:62}) to transform the first term.

\end{document}